\documentstyle[prd,twocolumn,eqsecnum,aps]{revtex}

\begin{document}

\draft
\title{Preheating of the nonminimally coupled inflaton field}
\author{Shinji Tsujikawa \thanks{electronic
address:shinji@gravity.phys.waseda.ac.jp}}
\address{ Department of Physics, Waseda University,
Shinjuku, Tokyo 169-8555, Japan\\[.3em]}
\author{Kei-ichi Maeda$^{1,2}$\thanks{electronic
address:maeda@gravity.phys.waseda.ac.jp}}
\address{ Department of Physics, Waseda University,
Shinjuku, Tokyo 169-8555, Japan\\[.3em]
Advanced Research Institute for Science and Engineering,\\
Waseda University, Shinjuku, Tokyo 169-8555, Japan\\[.3em]}
\author{Takashi Torii \thanks{electronic
address:torii@th.phys.titech.ac.jp}}
\address{Department of Physics, Tokyo
Institute of Technology, Meguro, Tokyo 152, Japan\\~}
\date{\today}
\maketitle
\begin{abstract}
We investigate preheating of an inflaton field $\phi$ coupled nonminimally to 
a spacetime curvature by the method of Hartree approximations.
In the case of a self-coupling inflaton potential $V(\phi)=\lambda
\phi^4/4$, the dynamics of preheating changes 
by the effect of the negative $\xi$.
We find that the nonminimal coupling works in two ways.
First, since the initial value of inflaton field for reheating 
becomes smaller with the increase of $|\xi|$, 
the evolution of the inflaton quanta is delayed
for fixed $\lambda$.
Second, the oscillation of the inflaton field is modified and the
nonadiabatic change around  $\phi=0$ occurs significantly.
That makes the resonant band of the fluctuation field wider.
Especially for strong coupling regimes $|\xi| \gg 1$,
the growth of the inflaton fluctuation is dominated by the 
resonance due to the nonminimal coupling,
which leads to the significant enhancement of low momentum modes.  
Although the final variance of the inflaton fluctuation does not change
significantly compared with the minimally coupled case, we have found that 
the energy transfer from the homogeneous inflaton to
created particles efficiently occurs for 
$\xi~\mbox{\raisebox{-1.ex}{$\stackrel
     {\textstyle<}{\textstyle \sim}$}}~-60$. 
\end{abstract}

\pacs{98.80.Cq, 05.70.Fh, 11.15.Kc}


%
\section{Introduction}                            %

Recently it has been recognized that a post-inflationary 
reheating begins by an explosive particle production 
which is called {\it preheating}\cite{TB,KLS1}.
As compared with the old theory of reheating\cite{DA}, 
preheating is a non-perturbative
process by which
the energy of the coherently oscillating inflaton field $\phi$ is 
efficiently transferred 
to scalar or other particles by parametric resonance.
The existence of this stage would drastically change the thermal 
history of the Universe, and 
many authors have investigated several issues such as baryogenesis
at the GUT scale\cite{baryogenesis}, the non-thermal phase transition
\cite{nonthermal}, the formation of topological defects 
\cite{defect}, and gravitational waves\cite{GW1,GW2}.

There are two typical models of preheating 
in the chaotic inflation scenario.
One of which is the massive inflaton $m^2\phi^2/2$ case.
In this model, another scalar field $\chi$ 
coupled to the inflaton field can be resonantly amplified.
There are several numerical works by making use of 
Hartree approximations\cite{KT1} and fully nonlinear lattice 
simulations\cite{KT2,PR}.
At the linear stage of preheating, the evolution of
produced fluctuations can be analyzed analytically by 
making use of the Mathieu equation\cite{KLS2}.
When the coupling constant is large, parametric resonance 
turns on from the broad resonance regime, 
and an efficient particle production
is possible overcoming the diluting effect by the expansion
of the Universe.
However, if the inflaton field $\phi$ does not couple to
another scalar field $\chi$, the inflaton fluctuation 
does not grow because the resonant terms are 
absent in a single field model.

Another model is the self-coupling inflaton 
$\lambda\phi^4/4$ case.
In this case, the fluctuation of the inflaton field 
is generated
even if we do not introduce another scalar field coupled to inflaton. 
The non-equilibrium scalar field dynamics are studied by fully
nonlinear calculations\cite{KT3}, and are also investigated
by making use of Hartree or large-$N$ approximations
\cite{Boy,Baa,Kaiser,Son,Greene}.
Since the Ricci scalar averaged over several oscillations of the inflaton 
field vanishes, the system can be
reduced to the theory of the Minkowski spacetime.
Making use of the properties of the Lam\'e equation, 
we can find that the growth rate of the inflaton quanta
is not as large as  that of the $\chi$ field in the 
 $m^2\phi^2/2+g^2\phi^2\chi^2/2$ model.
However, the inflaton fluctuation grows quasi-exponentially
and reaches  the maximum value $\langle\delta\phi^2\rangle
\sim 10^{-7} M_{\rm PL}^2$.
In both models, the backreaction effect on the oscillating
inflaton field plays an important role to shut off the resonance.
Also, a rescattering effect becomes significant at the final stage 
of preheating.

Generally, it has been assumed that a nonminimal coupling of 
the inflaton field or other scalar fields can be negligible in preheating 
phase. However, it was first pointed out in the massive 
inflaton model that the nonminimally coupled $\chi$ field 
can grow by the oscillating scalar curvature\cite{BL} 
especially when the coupling constant $\xi$ is negative.
Later we  investigated
the structure of resonance including the $\phi$-$\chi$ coupling 
where $\chi$ is nonminimally coupled to the scalar curvature~\cite{TMT1}.
In that model, resonance occurs in the wider range 
as compared with the minimally coupled
$m^2\phi^2/2+g^2\phi^2\chi^2/2$ model,
and the final variance of $\chi$-particles can be larger if 
$\xi~\mbox{\raisebox{-1.ex}{$\stackrel
     {\textstyle<}{\textstyle \sim}$}}~-4$.
In other inflation models such as the higher-curvature gravity, 
$\chi$-particles can be produced quite efficiently when $|\xi|$
is of order unity\cite{TMT2}.

By these works, we can expect that the nonminimal coupling 
modifies the structure of the resonance also in the inflation
model without other fields and makes preheating more efficient.
In the massive inflaton model, since $\xi$ is constrained to be
small as $|\xi|~\mbox{\raisebox{-1.ex}{$\stackrel
     {\textstyle<}{\textstyle \sim}$}}~10^{-3}$ by the observation of  
the Cosmic Background Explorer (COBE) satellite
\cite{FM}, the evolution of the system will not
change so much even if the nonminimal coupling is taken into 
account. 
On the other hand, in the $\lambda\phi^4/4$ model, 
the constraint of $\xi$ is absent for the negative coupling. 
Especially when $\xi$ is largely negative, the 
fine-tuning problem of $\lambda$ is relaxed\cite{FU,MS}.
It was pointed out in Ref.~$\cite{Boy,Kaiser}$ that 
the nonminimal coupling with the inflaton field
would not affect the evolution of the inflaton fluctuation
because the Ricci scalar averaged over 
several oscillations of the inflaton vanishes in reheating phase.
However, since the Ricci scalar oscillates 
by the interaction with  the homogeneous inflaton field, there is a possibility
that the dynamics of preheating would be altered
if we consider the nonminimal coupling.
In the minimally coupled case, it is known that the energy transfer 
from the homogeneous inflaton to produced particles is not efficient.
In the nonminimally coupled case, however, there is a possibility that 
enhanced momentum modes  become wider and the efficient particle
production is realized.
It is worth investigating the dynamics of the nonminimally 
coupled inflaton field, because this would affect the non-thermal 
phase transition\cite{nonthermal}, and the formation of topological defects 
\cite{defect}.

In this paper,  
we consider the dynamics of the nonminimally coupled inflaton
field in the chaotic inflation model.
We suppose that there are no fields other than inflaton. 
The Hartree approximation is used to estimate the variance of
produced fluctuations.
We organize our paper as follows.
In the next section, we present basic equations of the 
background and scalar fields.
In Sec.~III, we study the dynamics of the nonminimally coupled 
massless inflaton in the inflationary phase, and present the initial
values of the inflaton
when the reheating stage starts.  
In Sec.~IV,  we investigate how the nonminimal coupling changes
the evolution of the system in preheating phase.
The final variance of the fluctuation is also presented.
We give our conclusion and discussion in the final section.

\section{Basic equations}   

We investigate a model where an inflaton field $\phi$ is nonminimally 
coupled with a scalar curvature $R$:
\begin{eqnarray}
{\cal L} = \sqrt{-g} \left[ \frac{1}{2\kappa^2}R
   -\frac{1}{2}(\nabla \phi)^2
   -V(\phi)
   -\frac12 \xi R \phi^2
    \right],
\label{B1}
\end{eqnarray}
where $\kappa^{2}/8\pi \equiv G =M_{\rm PL}^{-2} $ is Newton's
gravitational constant, and $\xi$ is a coupling constant.
$V(\phi)$ is a potential of the inflaton field
which is described by
\begin{eqnarray}
V(\phi)=\frac12 m^2 \phi^2 +\frac14 \lambda \phi^4.
\label{B2}
\end{eqnarray}
In this paper, we mainly consider the self-coupling inflaton model.

In the minimally coupled case, 
the value of the inflaton field at the time $t_I$ when the inflation turns on
is required $\phi (t_I)~\mbox{\raisebox{-1.ex}{$\stackrel
     {\textstyle>}{\textstyle\sim}$}}~5M_{\rm PL}$ 
in order to solve the cosmological puzzles.
Also, the primordial density perturbation observed by the COBE 
satellite constrains the coupling of the inflaton field
as $m~\mbox{\raisebox{-1.ex}{$\stackrel
     {\textstyle<}{\textstyle \sim}$}}~10^{-6} M_{\rm PL}$ for the massive inflaton model,
and $\lambda~\mbox{\raisebox{-1.ex}{$\stackrel
     {\textstyle<}{\textstyle \sim}$}}~10^{-12}$ for the self-coupling inflaton model 
respectively.

However, the dynamics of the Universe changes
if the nonminimal coupling is taken into account.
Futamase and one of the present authors\cite{FM} showed 
that in the case of the positive coupling, $\xi$ is constrained to be small as 
 $\xi~\mbox{\raisebox{-1.ex}{$\stackrel
     {\textstyle<}{\textstyle \sim}$}}~10^{-3}$ in both chaotic models.
This is due to the constraint that the effective gravitational
constant
\begin{eqnarray}
G_{\rm eff}=\frac{G}{1-\phi^2/\phi^2_c},~~{\rm with}~~
\phi^2_c \equiv \frac{M^2_{\rm PL}}{8\pi\xi},
\label{B80}
\end{eqnarray}
must be positive to connect to our present universe, which
yields $\phi< \phi_c =M_{\rm PL}/\sqrt{8\pi\xi}$.
The requirement that $\phi_c>\phi (t_I)$ 
leads the constraint: $\xi~\mbox{\raisebox{-1.ex}{$\stackrel
     {\textstyle<}{\textstyle \sim}$}}~10^{-3}$.

The Lagrangian in Eq.~$(\ref{B1})$ is reduced to the system
in the Einstein frame by a conformal transformation
\begin{eqnarray}
\hat{g}_{\mu \nu}=\Omega^2 g_{\mu \nu},
\label{B3}
\end{eqnarray}
where $\Omega^2 \equiv 1-\xi\kappa^2\phi^2$.
Then we obtain the following equivalent Lagrangian:
\begin{eqnarray}
{\cal L} = \sqrt{-\hat{g}} \left[ \frac{1}{2\kappa^2} \hat{R}
   -\frac{1}{2} F^2 (\hat{\nabla} \phi)^2
   -\hat{V} (\phi) \right],
\label{B4}
\end{eqnarray}
where 
\begin{eqnarray}
F^2 \equiv \frac{1-(1-6\xi) \xi\kappa^2 \phi^2 }
{(1-\xi\kappa^2 \phi^2)^2},
\label{B5}
\end{eqnarray}
and
\begin{eqnarray}
\hat{V} (\phi)
\equiv \frac{V(\phi)}{(1-\xi\kappa^2\phi^2)^2}.
\label{B50}
\end{eqnarray}
Introducing a new scalar field $\psi$ as
\begin{eqnarray}
\psi \equiv \int F(\phi) d\phi,
\label{B6}
\end{eqnarray}
the Lagrangian in the new frame is reduced to the canonical form:
\begin{eqnarray}
{\cal L} = \sqrt{-\hat{g}} \left[ \frac{1}{2\kappa^2} \hat{R}
   -\frac{1}{2} (\hat{\nabla} \psi)^2
   -\hat{V} (\psi)  \right].
\label{B7}
\end{eqnarray}

When $\xi$ is negative, the effective potential $(\ref{B50})$ shows 
a different characteristic from the minimally coupled case.
In the massive inflaton case, since the potential has a maximum at 
$\phi=M_{\rm PL}/\sqrt{8\pi|\xi|}$ (See Fig.~1),
$\xi$ is constrained to be small as 
$|\xi|~\mbox{\raisebox{-1.ex}{$\stackrel
     {\textstyle<}{\textstyle \sim}$}}~10^{-3}$.
As long as we do not introduce another field 
coupled to inflaton, the fluctuation of the inflaton quanta is not generated
in the minimally coupled case.
Hence the situation does not change even if we take into account 
such small values of $|\xi|$.

However, such a constraint is absent in the case 
of the $\lambda\phi^4/4$ model, because the effective
potential of inflaton
has a plateau to  lead a sufficient inflation (Fig.~1).
Especially when the coupling is strong as 
$|\xi|~\mbox{\raisebox{-1.ex}{$\stackrel
     {\textstyle>}{\textstyle\sim}$}}~1$, several 
authors\cite{FU,MS,KF} investigated the dynamics 
of the nonminimally coupled inflaton field in the inflationary phase.
Hereafter, we consider the model of self-coupling potential
$V(\phi)=\lambda\phi^4/4$ with negative $\xi$.

Let us obtain the basic equations in the self-coupling model 
with the nonminimal coupling.
We can derive the equivalent equations in both frames 
of $(\ref{B1})$ and $(\ref{B7})$.
However, the equations in both frames are different each other 
if we take the mean field approximation in each frame.
Generally, it is convenient to consider the evolution of the system
in the Einstein frame.
However, since the conformal factor is included in the denominator
of the potential $(\ref{B50})$, we can not replace the mean field 
value of $\langle1/(1-\xi\kappa^2\phi^2)^2\rangle$ as 
$1/(1-\xi\kappa^2\langle\phi^2\rangle)^2$.
If these replacements were done, we need further assumptions 
of the derivative terms of the $\phi$ field in order to coincide with 
the basic equations of which the mean field approximation
is taken in the original frame.
Although we expect that these derivative terms do not affect the 
preheating dynamics significantly, there still remains the ambiguity
whether these assumptions are always valid. 
 
Moreover, the conformal factor $\Omega^2$ includes quantum 
variable $\phi^2$ since we consider the quantum 
fluctuation of the $\phi$ field.
The metric in the Einstein frame is perturbed 
by making a conformal transformation even if the metric in the original 
frame is spatially  homogeneous.
The analysis of other preheating models including the metric 
perturbation are performed by several authors\cite{mper}, 
but in the present model,
there remains the difficulty of how we take the mean field
approximation in the Einstein frame.
In order to avoid these problems which are accompanied by a conformal 
transformation, we consider the basic equations in the original 
frame and take the mean field approximation.

We adopt the flat Friedmann-Robertson-Walker metric 
as the background spacetime;
\begin{eqnarray}
ds^2 = -dt^2 + a^2(t) d {\bf x}^2,
\end{eqnarray}
where $a(t)$ is the scale factor.
Although we do not consider the metric perturbation in this paper, 
we should include this effect for a complete study of preheating. 
It is under consideration.

We obtain the following field equations 
from the Lagrangian $(\ref{B1})$,
\begin{eqnarray}
\frac{1-\xi\kappa^2\phi^2}{\kappa^2} G_{\mu\nu}
 &= & (1-2\xi)\nabla_{\mu} \phi \nabla_{\nu} \phi 
-\left(\frac12 -2\xi \right) g_{\mu \nu} (\nabla \phi)^2
\nonumber \\
 -g_{\mu \nu} V(\phi) &+&2\xi \phi (g_{\mu \nu} \kern1pt\vbox{\hrule height
1.2pt\hbox{\vrule width 1.2pt\hskip 3pt
   \vbox{\vskip 6pt}\hskip 3pt\vrule width 0.6pt}\hrule
height 0.6pt}\kern1pt -\nabla_{\mu}
\nabla_{\nu}) \phi,
\label{B9}
\end{eqnarray}
\begin{eqnarray}
\kern1pt\vbox{\hrule height
1.2pt\hbox{\vrule width 1.2pt\hskip 3pt
   \vbox{\vskip 6pt}\hskip 3pt\vrule width 0.6pt}\hrule
height 0.6pt}\kern1pt \phi -\xi R \phi -V,_{\phi}=0,
\label{B100}
\end{eqnarray}
where $\kern1pt\vbox{\hrule height
1.2pt\hbox{\vrule width 1.2pt\hskip 3pt
   \vbox{\vskip 6pt}\hskip 3pt\vrule width 0.6pt}\hrule
height 0.6pt}\kern1pt$ and $V,_{\phi}$ are defined as
 $\kern1pt\vbox{\hrule height
1.2pt\hbox{\vrule width 1.2pt\hskip 3pt
   \vbox{\vskip 6pt}\hskip 3pt\vrule width 0.6pt}\hrule
height 0.6pt}\kern1pt \equiv \partial_{\mu}
(\sqrt{-g} g^{\mu \nu} \partial_{\nu})/\sqrt{-g}$,  $V,_{\phi}
\equiv \partial V/\partial \phi$ respectively. 

Since we consider the quantum fluctuation of inflaton, 
the $\phi$ field is
represented by the homogeneous and fluctuational parts as
\begin{eqnarray}
\phi(t,{\bf x})=\phi_0(t) +\delta \phi(t,{\bf x}).
\label{B11}
\end{eqnarray}
As for the fluctuation part, we impose the tadpole condition
\begin{eqnarray}
\langle \delta \phi(t,{\bf x}) \rangle=0,
\label{B14}
\end{eqnarray}
where $\langle \cdots \rangle$ denotes the expectation value.
We also make the Hartree factorization as
\begin{eqnarray}
(\delta \phi)^3   \to   3 \langle (\delta \phi)^2 \rangle
(\delta \phi), 
\\
(\delta \phi)^4   \to  6 \langle (\delta \phi)^2 \rangle
(\delta \phi)^2-3\langle (\delta \phi)^2 \rangle^2.
\label{B13}
\end{eqnarray}
Then we obtain the Hamiltonian constraint equation by taking the 
mean field approximation in Eq.~$(\ref{B9})$,
\begin{eqnarray}
  H^2 &=&
   \frac{\kappa^2}{3(1-\alpha)}
     \Biggl[ \frac12 (\dot{\phi}_0^2+
    \langle \delta \dot{\phi}^2 \rangle)+
    \left( \frac12 -2\xi \right) \langle \delta\phi'^2 \rangle
    \nonumber \\
  &&  +\frac14 \lambda(\phi_0^4+6\phi_0^2 \langle \delta\phi^2 \rangle
    +3\langle \delta\phi^2 \rangle^2) 
    \nonumber \\ 
&& +2\xi \{3H(\phi_0 \dot{\phi}_0+\langle \delta\phi \delta
\dot{\phi} \rangle) -\langle \delta\phi \delta \phi'' \rangle \}
\Biggr],
\label{B70}
\end{eqnarray}
where a dot and a prime denote the derivative with respect to time
and space coordinates, respectively, and $\alpha \equiv \xi\kappa^2 \langle
\phi^2\rangle$, $H \equiv \dot{a}/a$.

Next, we consider the equations of the homogeneous and
fluctuational parts of the inflaton field.
Let us first obtain the expression of the scalar curvature $R$.
Since $R$ can be written by $R=-g^{\mu\nu} G_{\mu\nu}$,
we easily find the following relation from Eq.~$(\ref{B9})$ as
\begin{eqnarray}
\frac{1-\xi\kappa^2\phi^2}{\kappa^2} R
=(1-6\xi) (\nabla \phi)^2+4V(\phi)-6\xi\phi \kern1pt\vbox{\hrule height
1.2pt\hbox{\vrule width 1.2pt\hskip 3pt
   \vbox{\vskip 6pt}\hskip 3pt\vrule width 0.6pt}\hrule
height 0.6pt}\kern1pt \phi.
\label{B31}
\end{eqnarray}
Then $R$ is expressed as
\begin{eqnarray}
 R &=&
   \frac{\kappa^2}{1-\alpha}
     \Biggl[ (1-6\xi)(-\dot{\phi}_0^2 -\langle \delta\dot{\phi}^2 \rangle
    +\langle \delta\phi'^2 \rangle)
    \nonumber \\ 
&&    + \lambda(\phi_0^4+6\phi_0^2 \langle \delta\phi^2 \rangle
    +3\langle \delta\phi^2 \rangle^2) 
    +6\xi \{ \phi_0 \ddot{\phi}_0
 \nonumber \\ 
&& +\langle \delta\phi \delta
\ddot{\phi} \rangle+3H(\phi_0 \dot{\phi}_0+\langle \delta\phi \delta
\dot{\phi} \rangle) -\langle \delta\phi \delta \phi'' \rangle \}
\Biggr].
\label{B73}
\end{eqnarray}
Taking the mean field approximation of Eq.~$(\ref{B100})$, the 
$\phi_0$ and $\delta \phi$ fields obey the following equations of motion,
\begin{eqnarray}
\ddot{\phi}_0 +3H \dot{\phi}_0 +\lambda \phi_0
(\phi_0^2+3 \langle \delta\phi^2 \rangle)
+\xi R \phi_0 =0,
\label{B15}
\end{eqnarray}
\begin{eqnarray}
&&\delta \ddot{\phi} +3H \delta \dot{\phi}-\partial_i 
\partial^i (\delta\phi)
+\{ 3\lambda (\phi_0^2 + \langle \delta\phi^2 \rangle)
+\xi R \} \delta \phi=0.  
\nonumber \\
&& 
\label{B16}
\end{eqnarray}
Note that both evolutions of the $\phi_0$ and $\delta \phi$ fields
can be altered by the effect of the nonminimal coupling. 

In order to study a quantum particle creation,
we expand $\delta \phi$ as
\begin{eqnarray}
&&\delta \phi=\frac{1}{(2\pi)^{3/2}} \int 
\left(a_k 
\delta \phi_k(t)
 e^{-i {\bf k} \cdot {\bf x}}+a_k^{\dagger} 
\delta \phi_k^{*}(t)
 e^{i {\bf k} \cdot {\bf x}} 
\right) d^3{\bf k},
\nonumber \\
&  &
\label{B22}
\end{eqnarray}
where $a_k$ and $a_k^{\dagger}$ are the annihilation and creation
operators respectively.
Then, we find that $\delta \phi_k$ obeys the following equation
of motion
\begin{eqnarray}
&&\delta \ddot{\phi}_k +3H \delta \dot{\phi}_k
+\left[ \frac{k^2}{a^2}
+3\lambda (\phi_0^2 + \langle \delta\phi^2 \rangle)
+\xi R \right] \delta \phi_k=0.
\nonumber \\
& &
\label{B40}
\end{eqnarray}
The expectation values of $\delta \phi^2$ and $\phi^2$ are 
expressed as~\cite{Boy}
\begin{eqnarray}
\langle \delta \phi^2 \rangle &=& \frac1{2\pi^2} \int k^2
|\delta \phi_k|^2 dk,\\
\langle \phi^2 \rangle &=& \phi_0^2 
+\langle \delta \phi^2 \rangle.
\label{B27}
\end{eqnarray}

Introducing a conformal time $\eta \equiv \int a^{-1} dt$ and 
new scalar fields $\varphi_0 \equiv a \phi_0$ and  $\delta \varphi_k
\equiv a \delta \phi_k$, Eqs.~$(\ref{B15})$ and $(\ref{B16})$
are rewritten by
\begin{eqnarray}
&&\frac{d^2 \varphi_0}{d\eta^2}+\lambda \varphi_0 (\varphi_0^2+
3a^2  \langle \delta\phi^2 \rangle)
+\left( \xi R a^2 
-\frac{1}{a} \frac{d^2 a}{d\eta^2} \right) \varphi_0=0,
\nonumber \\
&&
\label{B28}
\end{eqnarray}
\begin{eqnarray}
\frac{d^2}{d\eta^2} \delta \varphi_k
+\omega_k^2 \delta \varphi_k=0,
\label{B29}
\end{eqnarray}
with
\begin{eqnarray}
\omega_k^2 \equiv
k^2 +3\lambda (\varphi_0^2+a^2 \langle \delta\phi^2 \rangle)
+\xi R a^2 -\frac{1}{a} \frac{d^2 a}{d\eta^2}.
\label{B60}
\end{eqnarray}
In the case of the minimal coupling, the $3\lambda\varphi_0^2$ term
in the time dependent frequency $(\ref{B60})$ causes the resonant 
growth of the inflaton fluctuation. In the nonminimally coupled case,
$\xi R a^2$ term also contributes to the parametric resonance.
In this paper, we mainly investigate how the inflaton fluctuation 
$\langle\delta \phi^2 \rangle$ grows by taking into account
the effect of the nonminimal coupling.

As for the initial conditions of the fluctuation, we choose the 
state that corresponds to the conformal vacuum as
\begin{eqnarray}
\delta \varphi_k(0)=\frac{1}{\sqrt{2\omega_k (0)}}, 
\\
\delta \dot{\varphi}_k(0)=-i \omega_k(0) \delta \varphi_k(0).
\label{B35}
\end{eqnarray}
We investigate the evolution of the 
inflaton quanta with those initial conditions as a semiclassical problem.
Before proceeding the particle creation in preheating phase,
we summarize the evolution of the scale factor
and the inflaton field in the inflationary period in the case 
of the strong coupling in the next section.

\section{The dynamics of inflation with the nonminimal coupling}    %
In this section, we analyze how inflation takes place in the presence
of the nonminimal coupling term. In the inflationary stage, since the 
fluctuation of the inflaton field is small compared with the 
homogeneous part,
we can approximately write 
Eqs.~(2.17), (2.18), and $(\ref{B15})$ as follows
\begin{eqnarray}
H^2 = \frac{\kappa^2}{3(1-\alpha)}
     \left[ \frac12 \dot{\phi}_0^2 
    +\frac14 \lambda \phi_0^4
    +6\xi H \phi_0 \dot{\phi}_0 \right],
\label{I1}
\end{eqnarray}
\begin{eqnarray}
R = \frac{\kappa^2}{1-\alpha}
     \left[ -(1-6\xi) \dot{\phi}_0^2 
    + \lambda \phi_0^4
   + 6\xi \phi_0 ( \ddot{\phi}_0+3H  \dot{\phi}_0)
    \right],  
\label{I2}
\end{eqnarray}
\begin{eqnarray}
\ddot{\phi}_0 +3H \dot{\phi}_0 +\lambda
\phi_0^3 +\xi R \phi_0  = 0,
\label{I3}
\end{eqnarray}
where $\alpha = \xi\kappa^2\phi_0^2$.
By Eq.~$(\ref{I2})$, the scalar curvature is eliminated 
to give
\begin{eqnarray}
\ddot{\phi}_0 +3H \dot{\phi}_0 +
\frac{\lambda\phi_0^3}{1-(1-6\xi)\alpha}-
\frac{\xi\kappa^2(1-6\xi)\dot{\phi}_0^2}
{1-(1-6\xi)\alpha}\phi_0 = 0.
\label{I4}
\end{eqnarray}
Imposing the slow roll approximations $|\ddot{\phi}_0/\dot{\phi}_0| \ll H$, 
$|\dot{\phi}_0/\phi_0| \ll H$, and $\frac12 \dot{\phi}_0^2 \ll V(\phi)$, 
Eqs.~$(\ref{I1})$ and  $(\ref{I4})$ can be written as
\begin{eqnarray}
H^2 = \frac{\kappa^2 \lambda \phi_0^4}{12(1-\alpha)}
\left[ 1-\frac{8\xi}{1-(1-6\xi)\alpha} \right],
\label{I5}
\end{eqnarray}
\begin{eqnarray}
3H\dot{\phi}_0 = -\frac{\lambda\phi_0^3}
{1-(1-6\xi)\alpha}.
\label{I6}
\end{eqnarray}

In order to see the effect of the nonminimal coupling, we consider the 
strong negative coupling case $\alpha \ll -1$. 
Then Eqs.~$(\ref{I5})$ and $(\ref{I6})$ are simplified as
\begin{eqnarray}
H^2 =-\frac{\kappa^2 \lambda \phi_0^4}{12\alpha},
\label{I7}
\end{eqnarray}
\begin{eqnarray}
3H\dot{\phi}_0 = \frac{\lambda\phi_0^3}
{(1-6\xi)\alpha}.
\label{I8}
\end{eqnarray}
Combining Eqs.~$(\ref{I7})$ and $(\ref{I8})$, we find the 
following relation
\begin{eqnarray}
\dot{\alpha} = -\frac{8\xi}{1-6\xi} H.
\label{I9}
\end{eqnarray}
This is easily integrated to give
\begin{eqnarray}
\alpha = \alpha_I - \frac{8\xi}{1-6\xi}
{\rm ln} \left( \frac{a}{a_I} \right),
\label{I10}
\end{eqnarray}
where the subscript $I$ denotes the values when the 
inflationary period starts.
We find by this relation that $|\alpha|$ decreases
with the expansion of the Universe. 
Since the scale factor satisfies the relation
\begin{eqnarray}
\frac{\dot{a}}{a}=
\sqrt{\frac{-\lambda \alpha_I}{12\xi^2\kappa^2}}
\left[ 1-\frac{8\xi}{(1-6\xi)\alpha_I} {\rm ln}
\left(\frac{a}{a_I} \right) \right]^{1/2},
\label{I11}
\end{eqnarray}
we find that $a$ is expressed as
\begin{eqnarray}
a=a_I \exp \left[ H_I t-\gamma (H_I t)^2 \right],
\label{I12}
\end{eqnarray}
where 
\begin{eqnarray}
H_I \equiv \sqrt{\frac{-\lambda \alpha_I}
{12\xi^2\kappa^2}},~~~\gamma \equiv \frac{2\xi}
{(1-6\xi)\alpha_I}.
\label{I13}
\end{eqnarray}
At the first stage of inflation, since the $\gamma$
term is negligible, the Universe evolves exponentially.
With the passage of time, however, $\gamma$ term
becomes important and the expansion rate of 
the Universe slows down.
Fakir and Unruh\cite{FU} showed that the inflationary period ends 
when $|\alpha|$ drops down to unity.
This can be easily seen that the term
\begin{eqnarray}
\left| \frac{\dot{\phi}}{H \phi} \right|
=\left| \frac{\dot{\alpha}}{2H \alpha} \right|
=\left| \frac{4\xi}{(1-6\xi)\alpha} \right|,
\label{I14}
\end{eqnarray}
becomes of order unity when $|\alpha| \approx 1$ in the case of
$\xi~\mbox{\raisebox{-1.ex}{$\stackrel
     {\textstyle<}{\textstyle \sim}$}}~-1$.

This criterion for the end of the inflation can also be obtained  
from the effective potential ($\ref{B50}$) in the Einstein frame.
The slow-roll parameter of the first order in the Einstein 
frame is defined by the potential $\hat{V}(\psi)$ as
\begin{eqnarray}
\epsilon \equiv \frac{1}{2\kappa^2}
\left(\frac{\hat{V}'(\psi)}{\hat{V}(\psi)}\right)^2.
\label{I15}
\end{eqnarray}
Making use of Eqs.~($\ref{B5}$)-($\ref{B6}$),
this can be written by 
\begin{eqnarray}
\epsilon =\frac{8}{\kappa^2\phi_0^2 [1-(1-6\xi)\xi\kappa^2\phi_0^2]}.
\label{I16}
\end{eqnarray}
Setting $\epsilon =1$, the value of $\alpha$ when the inflationary period
ends is estimated by
\begin{eqnarray}
|\alpha_F|=\frac{\sqrt{(1-24\xi)(1-8\xi)}-1}{2(1-6\xi)},
\label{I17}
\end{eqnarray}
where the subscript $F$ denotes the value at the time
$t_F$ when the inflation ends.
Note that $|\alpha_F|=1$ for $\xi=-1$. 
$|\alpha_F|$ slowly increases with the decrease of $\xi$ for 
$\xi<0$. However, since the limiting value for $\xi \to -\infty$
is  $|\alpha_F| =1.15$, inflation ends at $|\alpha_F| \approx 1$ 
for $\xi~\mbox{\raisebox{-1.ex}{$\stackrel
     {\textstyle<}{\textstyle \sim}$}}~-1$.

As for the final values of $\phi_0$ for the end of inflation,
we make use of the ones which are determined by Eq.~($\ref{I17}$) as
\begin{eqnarray}
\phi_0 (t_F)= \left[\frac{\sqrt{(1-24\xi)(1-8\xi)}-1}
{16\pi(1-6\xi)|\xi|} \right]^{1/2} M_{\rm PL}.
\label{I18}
\end{eqnarray}
In the case of the minimal coupling, $\phi_0 (t_F)=0.56 M_{\rm PL}$.
With the increase of $|\xi|$, $\phi_0 (t_F)$ decreases.
We show the values of $|\alpha_F|$ and $\phi_0 (t_F)$ 
in various cases in TABLE I. Since the large $|\xi|$ makes the inflaton
potential flatter, inflation takes place for rather small values of
$\phi_0$ as compared with the minimal coupling case.
As for the time $t_i$ at the start of reheating,
we assume $t_i=t_F$.
Hence, the initial value of the inflaton field for the preheating
is $\phi_0 (t_i)=\phi_0 (t_F)$ estimated by
Eq.~($\ref{I18}$).
In the next section we 
investigate the preheating dynamics of the 
nonminimally coupled inflaton field. 

\section{Preheating of the nonminimally coupled inflaton field }    %
In this section, we briefly review the minimally coupled inflaton case 
, and after that, investigate the case that the nonminimal coupling
is included.   

\subsubsection{$\xi=0$ case}   %

In the minimally coupled case, the evolution of the inflaton quanta 
can be obtained analytically by using the stability and instability 
chart of the Lam\'e equation.
First, making use of the time-averaged relation
$\frac12 \langle \dot{\phi}^2 \rangle_T =2 \langle V(\phi) \rangle_T$,
the evolution of the scale factor is approximately written as
$a \sim \eta$. Since the scalar curvature is expressed as
$R=\frac{6}{a^3} \frac{d^2 a}{d \eta^2}$, the time-averaged value 
of $R$ vanishes in the case of the minimal coupling.

Introducing a dimensionless time variable
$x \equiv \sqrt{\lambda} M_{\rm PL} \eta$
and scalar fields $f \equiv \varphi_0/M_{\rm PL}$,
$\langle \delta \bar{\phi}^2 \rangle \equiv 
\langle \delta \phi^2 \rangle/M_{\rm PL}^2$,
$\langle \delta \bar{\varphi}^2 \rangle \equiv 
a^2 \langle \delta \bar{\phi}^2 \rangle$,
Eq.~($\ref{B28}$) is reduced to the equation 
for the theory $\lambda \phi^4/4$ in Minkowski spacetime:
\begin{eqnarray}
\frac{d^2 f}{dx^2}+f^3
\left(1+3 \frac{\langle \delta \bar{\varphi}^2 \rangle}
{f^2} \right) =0.
\label{S2}
\end{eqnarray}
At the stage when the back reaction effect is negligible,
the solution of Eq.~$(\ref{S2})$ can be handled analytically.
Integrating the equation $d^2 f/dx^2+f^3=0$, we find 
\begin{eqnarray}
\left(\frac{df}{dx}\right)^2=
-\frac12 \left(f^4-f_{\rm co}^4 \right),
\label{S20}
\end{eqnarray}
where $f_{\rm co}$ is the value when the $\varphi_0$ field begins 
to oscillate coherently.
The solution of this equation is expressed by the elliptic cosine,
\begin{eqnarray}
f= f_{\rm co}~{\rm cn} \left(f_{\rm co} (x- x_{\rm co}) ; 
\frac{1}{\sqrt{2}} \right).
\label{S30}
\end{eqnarray}
This function is well approximated as
\begin{eqnarray}
f & \approx & f_{\rm co} \cos (0.8472f_{\rm co} x)
\nonumber \\
& = & f_{\rm co} \cos
 (0.8472 \sqrt{\lambda} \varphi_0(t_{\rm co})\eta),
\label{S21}
\end{eqnarray}
where we dropped the initial value of
$x_{\rm co}$ for simplicity.
Note that the frequency of the oscillation depends on the values of
$\lambda$ and $\varphi_0(t_{\rm co})$. 
As we have showed, initial values of the inflaton for preheating  
changes if we take into account the effect of the nonminimal coupling. 
We discuss this issue in the next section.
 
Since the inflaton field oscillates periodically, the inflaton quanta 
can be produced by parametric resonance.
Eq.~($\ref{B29}$)  can be written as
\begin{eqnarray}
\frac{d^2}{dx^2} \delta \varphi_k+
\left[ p^2 +3f^2 \left(1+
\frac{\langle \delta \bar{\varphi}^2 \rangle}
{f^2} \right) \right] \delta \varphi_k=0,
\label{S3}
\end{eqnarray}
where $p^2 \equiv k^2/(\lambda M_{\rm PL}^2)$ is the 
normalized momentum.
Eq.~($\ref{S3}$) belongs to the class of Lam\'e equation when the 
inflaton quanta
are not produced significantly.
The structures of resonance are studied precisely in Ref.~
\cite{Kaiser,Greene}.
Making use of the chart of stability and instability bands
of the Lam\'e equation, the growing mode of
$\delta \varphi_k$ exists in the narrow range:
\begin{eqnarray}
\frac32 \tilde{f}^2<
p^2<\sqrt{3} \tilde{f}^2,
\label{S4}
\end{eqnarray}
where variables with a tilde $\tilde{}$ means the amplitude of the 
variables in this paper.

As compared with the model of the massive inflaton plus another
scalar field $\chi$, we found that the modes whose momenta are
close to $k=0$ are not enhanced
and that the maximum growth rate of $\delta \varphi_k$ 
is not so large.
In spite of the small growth rate, $\langle \delta \phi^2\rangle$ grows 
quasi-exponentially by parametric resonance.
We show in Fig.~2 the evolution of $\langle \delta \phi^2\rangle$
for the initial value of $\phi_0 (t_i)=0.56 M_{\rm PL}$, and
the coupling $\lambda=10^{-12}$ which is constrained by
the COBE data\cite{com}.
At the initial stage, $\langle \delta \phi^2\rangle$ decreases
because the expansion rate of the Universe surpasses the particle
creation rate.
Our numerical calculation based on the mean field approximation
reveals that $\langle \delta \phi^2\rangle$ begins 
to increase after $x~\mbox{\raisebox{-1.ex}{$\stackrel
  {\textstyle>}{\textstyle\sim}$}}~100$, and reaches the maximum value,
\begin{eqnarray}
\langle \delta \phi^2 \rangle_f =
1.5 \times 10^{-7} M_{\rm PL}^2.
\label{S5}
\end{eqnarray}
The subscript $f$ denotes the value at the time $t_f$ when the
$\langle \delta \phi^2 \rangle$ reaches the maximum value.
This result agrees well with the fully non-linear calculation
performed by Khlebnikov and Tkachev\cite{KT3}.
Although our mean field approximation does not include
the rescattering effect (i.e.~the mode-mode coupling
of the inflaton fluctuation), the final variance is consistent 
with the lattice simulation.
This implies that the back reaction effect due to the  growth 
of the inflaton quanta plays an important role for the 
termination of parametric resonance.
As the inflaton fluctuation grows, the back reaction effect 
prevents the coherent oscillation of the $\varphi_0$ field as is found 
by Eq.~($\ref{S2}$).
The resonance band in Eq.~$(\ref{S3})$ is shifted away 
from the original position as particles are produced.
According to the analytic estimation 
by Greene et al.\cite{Greene},
the growth of the inflaton quanta stops 
when $\langle \delta \phi^2 \rangle$ grows up to
\begin{eqnarray}
\langle \delta \phi^2 \rangle_f  \approx
0.05 \tilde{\phi}_0^2.
\label{S6}
\end{eqnarray}
Our numerical value of the termination point 
is about $x_f \approx 440$ and $\tilde{\phi}_0  (x_f)\approx
1.4 \times 10^{-3} M_{\rm PL}$, which yields
$\langle \delta \phi^2 \rangle_f  \approx 9.8 \times 10^{-8}
M_{\rm PL}^2$ by Eq.~($\ref{S6})$.
This agrees well with the numerical value of  Eq.~$(\ref{S5})$.
The final fluctuation in the self-coupling inflation model is
rather small compared with other preheating models
as $m^2\phi^2/2+g^2\phi^2\chi^2/2$ or $m^2\phi^2/2+
\xi R \chi^2/2$\cite{KLS2,TMT1}.
In the massive inflaton model, preheating starts from the broad 
resonance regime in certain ranges of the coupling constant. 
Hence in this case, the growth rate of the fluctuation can 
take larger values, and the modes which are close to
$k=0$ are mainly enhanced.
In the present model, however, the instability bands are 
restricted as Eq.~$(\ref{S4})$, and parametric resonance
is not so efficient.
Parametric resonance ends when the dispersion of
the fluctuation becomes about 20\% of the amplitude 
of the inflaton field.
We mainly concern that how the effect of the nonminimal 
coupling would change the growth of the fluctuation and 
the final variance.
In the next subsection, we investigate this issue precisely.

\subsubsection{$\xi \ne 0$ case}   %

Next we proceed to 
the nonminimally coupled case. Its effect can be seen in several
places and changes the dynamics of preheating.
As we have already mentioned, 
the nonminimal coupling alters the initial
value of the inflaton field $\phi_0$ when the 
reheating period starts. With the increase of $|\xi|$, this initial value
decreases and the frequency of the inflaton oscillation becomes smaller.
Second, as is found by Eqs.~$(\ref{B15})$ and $(\ref{B16})$, 
the nonminimal 
coupling affects the evolutions of the $\phi_0$ and 
$\delta \phi_k$ fields through the oscillation of the 
scalar curvature.
At the stage before the inflaton quanta do not grow significantly,
the scalar curvature is approximately written by
\begin{eqnarray}
 R  \approx
\frac{\kappa^2 (1-6\xi) (\lambda \phi_0^4-\dot{\phi}_0^2)}
{1-(1-6\xi)\alpha},
\label{S7}
\end{eqnarray}
where we used Eqs.~$(\ref{I2})$ and $(\ref{I3})$.
Note that the oscillation of the scalar curvature depends on $\xi$
significantly. Hereafter, we investigate how the nonminimal coupling
would have influence on the evolution of $\phi_0$ and 
$\delta \phi_k$ fields.
We can rewrite Eqs.~$(\ref{B28})$ and $(\ref{B29})$ by taking 
notice of the relation $R=\frac{6}{a^3} \frac{d^2 a}{d \eta^2}$ as
\begin{eqnarray}
\frac{d^2 f}{dx^2}+f^3
\left(1+3 \frac{\langle \delta \bar{\varphi}^2 \rangle}
{f^2} \right) +
\left(\xi-\frac16 \right) a^2 \bar{R} f=0,
\label{S8}
\end{eqnarray}
\begin{eqnarray}
\frac{d^2}{dx^2} \delta \varphi_k+
\left[ p^2 + 3f^2 \left(1+
\frac{\langle \delta \bar{\varphi}^2 \rangle}{f^2} 
\right) \right.
\nonumber \\
+ \left. \left(\xi-\frac16 \right) a^2 \bar{R} \right] 
\delta \varphi_k=0,
\label{S9}
\end{eqnarray}
where $\bar{R} \equiv R/(\lambda M_{\rm PL}^2)$.
In the conformal coupling case ($\xi=1/6$), although it
breaks the constraint $\xi~\mbox{\raisebox{-1.ex}{$\stackrel
     {\textstyle<}{\textstyle \sim}$}}~10^{-3}$, the system is reduced
exactly to the equations in Minkowski spacetime.

In the case of the minimal coupling, the last terms in Eqs.~$(\ref{S8})$
and $(\ref{S9})$ become negligible after several oscillations 
of the $\phi_0$ field.
In the positive $\xi$ case, $\xi$ is constrained to be 
$\xi~\mbox{\raisebox{-1.ex}{$\stackrel
     {\textstyle<}{\textstyle \sim}$}}~10^{-3}$ for the inflationary period to proceed well.
We have numerically checked that the evolution of the system
is similar to that in the minimally coupled case.
Hereafter, we concentrate on the case when $\xi$ is 
negative and fix the self coupling constant as $\lambda=10^{-12}$.
At the end of this section, we will briefly comment the case 
when $\lambda$ is not fixed.

In the case of $-0.1~\mbox{\raisebox{-1.ex}{$\stackrel
     {\textstyle<}{\textstyle \sim}$}}~\xi <0$, since the initial value of $|\alpha|$
is much smaller than unity, the effect of the nonminimal coupling does 
not alter the evolution of the inflaton significantly compared with 
the minimally coupled case.
When $-10~\mbox{\raisebox{-1.ex}{$\stackrel
     {\textstyle<}{\textstyle \sim}$}}~\xi~\mbox{\raisebox{-1.ex}{$\stackrel
     {\textstyle<}{\textstyle \sim}$}}~-0.1$,  $\xi$ effects begin to be important
and change the evolution of the system.
In this case, it appears mainly in
the value of $\phi_0 (t_i)$ and  $\phi_0 (t_{\rm co})$. 
Let us consider the case of $\xi=-1$ as an example.
In this case, preheating begins when 
$\phi_0(t_i)=0.1995 M_{\rm PL}$ and $|\alpha_i|=1$.
Since the value of $\phi_0 (t_{\rm co})$ 
is small compared with the minimally coupled case,
the period of the oscillation of the inflaton field 
becomes about five times longer than the $\xi =0$ case (See Fig.~3).
The growth of the inflaton quanta is delayed due to this effect.
Since $|\alpha|$ is of order unity at the end of inflation, 
we expect that the evolution of the inflaton field and
the inflaton quanta would be 
affected by the presence of the $(\xi-\frac16)a^2 \bar{R}$ 
terms in Eqs.~$(\ref{S8})$
and $(\ref{S9})$.
However, after several oscillations of the inflaton field,
$\phi_0$ drops down under $0.02 M_{\rm PL}$, which means that 
$|\alpha|$ rapidly becomes much smaller than unity.
We have numerically checked that these terms are negligible
compared with the  $f^3 \left(1+3 \frac{\langle \delta \bar{\varphi}^2 \rangle
}
{f^2} \right)$ and $3f^2 \left(1+\frac{\langle \delta \bar{\varphi}^2 
\rangle}{f^2} \right)$ terms for 
$x~\mbox{\raisebox{-1.ex}{$\stackrel
     {\textstyle>}{\textstyle\sim}$}}~100$.
Hence the structure of resonance for 
$x~\mbox{\raisebox{-1.ex}{$\stackrel
     {\textstyle>}{\textstyle\sim}$}}~100$ is almost 
the same as that in the minimally coupled case.

Since the particle creation rate is smaller than the expansion rate 
of the Universe at the initial stage of preheating, 
$\langle\delta \phi^2 \rangle$ 
decreases at the first stage (See Fig.~2).
The numerical calculation shows that $\langle\delta \phi^2 \rangle$
begins to grow after $x \approx 400$, which means that 
the evolution of the inflaton quanta is not affected by the $\xi$ effect
directly.
Although the growth of $\langle\delta \phi^2 \rangle$ is delayed 
due to the decrease of the initial value of the inflaton field, 
the final variance is almost the same as the $\xi=0$ case.
The last term in Eq.~$(\ref{S9})$ is completely negligible after
$\langle \delta\phi^2 \rangle$ reaches to its maximum value, and 
the final fluctuation can be analyzed in the same way as the minimally 
coupled case.  Numerical values of the maximum point 
$\langle\delta \phi^2 \rangle_f=1.7 \times 10^{-7} M_{\rm PL}^2$
and $\tilde{\phi}_0 (x_f)=1.4 \times 10^{-3}$ are almost the same 
as those in the minimally coupled case,
and parametric resonance terminates 
before the energy of the $\phi_0$ field is sufficiently
transferred to the inflaton fluctuation.
In this case, the enhanced mode of the momentum exists also 
in the narrow range which is estimated by Eq.~$(\ref{S4})$.

For $\xi~\mbox{\raisebox{-1.ex}{$\stackrel
     {\textstyle<}{\textstyle \sim}$}}~-10$, the nonminimal coupling becomes so strong
that the structure of the resonance changes.
Let us first consider the dynamics of preheating
in the strong coupling limit: $|\xi| \gg 1$, and later we show
the concrete cases.

We assume that the amplitude of the inflaton field is not very small, i.e.,
$|\xi| \gg \tilde{\alpha}\xi \gg 
1$ at the initial stage of preheating, where $\tilde{\alpha}=
\xi \kappa^2 {\tilde{\phi}_0}^2$.
When the $\phi_0$ field oscillates far from the minimum of the 
potential ($\phi_0 \approx \tilde{\phi}_0$),
we can approximately express the scalar curvature $(\ref{S7})$ as 
\begin{eqnarray}
\bar{R} \approx \frac{8\pi}{\alpha} 
\left(1-\frac{1}{6\xi\alpha} \right)  \left[
-\bar{\phi}_0^4 +\frac{1}{a^2} \left( \frac{d\bar{\phi}_0}{dx}
\right)^2 \right],
\label{S10}
\end{eqnarray}
where $\bar{\phi}_0 \equiv \phi_0/M_{\rm PL}$.
Here we did not neglect the 
term $1/(6\xi\alpha)$ in Eq.~$(\ref{S10})$. 
This term increases with the passage of time because $|\tilde{\alpha}|$ decreases
from the order of unity as long as $\langle\delta\phi^2 \rangle \ll 
\tilde{\phi}_0^2$.
We find from Eq.~$(\ref{S8})$ that the equation of the $\phi_0$
field is written by
\begin{eqnarray}
\frac{d^2 f}{dx^2} +\frac{1}{6\xi} \left( \frac{1}{\alpha}
+1 \right) f^3 \approx 0.
\label{S11}
\end{eqnarray}
Note that the changing rate of the $f$ field becomes very small 
around $\phi_0 \approx \tilde{\phi_0}$
in the case of the strong coupling. 
This behavior is clearly seen in Fig.~4, which shows 
the evolution of the $\phi$ field in the $\xi=-100$ case.

During one oscillation of 
the inflaton field,  $|(1-6\xi)\alpha|$  becomes small 
around the minimum of the potential.
In this case, the scalar curvature is approximated as
\begin{eqnarray}
\bar{R} \approx 48\pi \xi \left[
-\bar{\phi}_0^4 +\frac{1}{a^2} \left( \frac{d\bar{\phi}_0}{dx}
\right)^2 \right].
\label{S13}
\end{eqnarray}
Then the $\phi_0$ field satisfies the following equation of motion
\begin{eqnarray}
\frac{d^2 f}{dx^2} & +&\left[1-48\pi\xi^2
\left\{ \bar{\phi}_0^2-\frac{1}{a^2} \left( \frac{d\bar{\phi}_0}{dx}
\right)^2  \right\} \right] f^3 \approx 0.
\label{S14}
\end{eqnarray}
In the very vicinity of $\phi_0 =0$, Eq.~$(\ref{S14})$ yields
\begin{eqnarray}
\frac{d^2 f}{dx^2}+ \left[1+48\pi\xi^2
\frac{1}{a^2} \left( \frac{d\bar{\phi}_0}{dx} \right)^2 \right]
f^3 \approx 0.
\label{S15}
\end{eqnarray}
Because of the existence of the last term, the oscillation becomes rapid 
compared with the minimally coupled case (See Fig.~4).
During one oscillation of the inflaton field, the equation changes from 
$(\ref{S11})$ to $(\ref{S14})$, and  from $(\ref{S14})$ to $(\ref{S11})$
as long as the amplitude of the $\phi_0$ field satisfies 
the condition $\tilde{\alpha} \xi \gg 1$.
With the passage of time, however, since the amplitude 
$\tilde{\phi}_0$ decreases by the expansion of the universe,
the equation of the $\phi_0$ field approaches to that
of the minimally coupled case.

Next, let us consider the fluctuation of the inflaton field.
The particle production occurs when the frequency $(\ref{B60})$ 
changes nonadiabatically~\cite{KLS2}.
Since the $\phi_0$ field rapidly varies around $\phi_0 \approx 0$
for the strong coupling $|\xi| \gg 1$, the fluctuation grows 
nonadiabatically when $\phi_0$ is passing through zero.
Substituting the relation $(\ref{S13})$ into $(\ref{S9})$,
the equation of the fluctuation around $\phi_0 \approx 0$ is approximately
written by\cite{comment}
\begin{eqnarray}
\frac{d^2}{dx^2} \delta \varphi_k &+&
\Biggl[ p^2 +3f^2 
-48\pi a^2 \xi^2 \biggl\{ \bar{\phi}_0^4
\nonumber \\
 &-& \frac{1}{a^2} \left( \frac{d\bar{\phi}_0}{dx}
\right)^2 \biggr\} \Biggr]
\delta \varphi_k \approx 0.
\label{S17}
\end{eqnarray}
In the vicinity of $\phi_0=0$, the 
$48\pi\xi^2\left( \frac{d\bar{\phi}_0}{dx} \right)^2$ 
term dominates over other terms.
We have numerically checked that this term changes most 
significantly around $\phi_0 \approx 0$ and mainly causes
the nonadiabatic increase of the fluctuation.
The structure of resonance is modified by this effect, and the range 
of the enhanced momentum modes becomes broader.
As was  presented by Greene et al.\cite{Greene}, 
the minimally coupled case (i.e. the prefactor of $f^2$ is 3) 
is the least favorable case for an efficient preheating. 
In the present model, however, the effect of the nonminimal
coupling makes the resonance band wider and we can expect 
the rapid growth of the fluctuation. 

Let us consider the case of $\xi=-20$. In this case, initial
values for preheating are $\phi_0(t_i)=0.0477 M_{\rm PL}$ and 
$|\alpha_i|=1.145$ (See Table I).
At the initial stage of preheating, since the structure of resonance
is dominated by Eq.~$(\ref{S17})$ around $\phi_0=0$
where the particle production occurs,
momentum modes close to $k=0$ can be enhanced. 
In Fig.~5 we show the evolution of $\omega_1 \equiv 
3f^2 \left(1+ \frac{\langle \delta \bar{\varphi}^2 \rangle}
{f^2} \right)$ and 
$\omega_2 \equiv \left(\xi-\frac16 \right) a^2 \bar{R}$ 
terms in Eq.~$(\ref{S9})$.
We find that the nonminimally coupled term
$\omega_2$ has larger amplitude than $\omega_1$ 
at the initial stage. 
Since the periods of both functions are the same,
$\omega_2$ dominates over $\omega_1$ for 
the nonadiabatic change around $\phi_0=0$.
However, this term becomes negligible 
with the decrease of the scalar curvature.
The increase of $\langle \delta\phi^2 \rangle$ occurs 
from $x \approx 4000$, because the particle creation rate can not 
surpass the expansion rate of the Universe for
$x~\mbox{\raisebox{-1.ex}{$\stackrel
     {\textstyle<}{\textstyle \sim}$}}~4000$.
Since the contribution of $\omega_2$ is about 10\% of 
$\omega_1$ at $x \approx 4000$, momentum modes 
close to $k=0$ are marginally enhanced when 
$\langle \delta\phi^2 \rangle$ begins to increase.
After $x~\approx~6000$, this contribution becomes less 
than 5\% and the structure of resonance approaches to 
the minimally coupled case.
Actually we can see in Fig.~6 that the growth rate of 
$\langle \delta\phi^2 \rangle$ becomes a bit small for 
$x~\mbox{\raisebox{-1.ex}{$\stackrel
     {\textstyle>}{\textstyle\sim}$}}~6000$.
Although the range of enhanced $k$ modes is wide at the 
initial stage of preheating, it becomes narrow gradually as preheating
proceeds and only the momentum modes which satisfy
the relation $(\ref{S4})$ are enhanced.
After $x \approx 6000$, the fluctuation increases up to 
$\langle \delta\phi^2 \rangle_f=1.3 \times 10^{-7}M_{\rm PL}^2$,
(at $x_f=1.04 \times 10^4$), which is almost the same 
as the $\xi=0$ case.
Although the resonant structure is different at the initial stage,
this does not affect the final abundance of the fluctuation significantly.
One important property is that the final amplitude of the 
$\phi_0$ field is $\tilde{\phi}_0(x_f)=8.0 \times 10^{-4}$, 
which yields $\langle \delta\phi^2\rangle_f  \approx 0.2 \tilde{\phi}_0^2$.
This means that the energy of the homogeneous inflaton field is more 
efficiently transferred to the inflaton fluctuation than in the 
minimally coupled case.

With the increase of $|\xi|$, the duration of the initial resonant stage 
caused by the nonminimal coupling continues longer.
However, as time passes and $|\alpha|$ decreases, the structure of 
resonance approaches  the minimally coupled case.
In the case of $\xi=-50$, the initial value of the inflaton is 
$\phi_0 (t_i)=0.0303 M_{\rm PL}$, which is much smaller than that of
 the $\xi=0$ case.
Because of this, the growth of the inflaton fluctuation is delayed.
At the initial stage, low momentum modes can be enhanced, and this is 
much more efficient than the minimally coupled case with the same 
initial value of the inflaton. 
However, for $4000~\mbox{\raisebox{-1.ex}{$\stackrel
     {\textstyle<}{\textstyle \sim}$}}~x~\mbox{\raisebox{-1.ex}{$\stackrel
     {\textstyle<}{\textstyle \sim}$}}~7000$, $\langle\delta \phi^2 \rangle$ 
shows a short plateau.
This means that the range of enhanced $k$ modes becomes narrow 
after $x \approx 4000$, and the contribution of low momentum modes
becomes less important.
For $x~\mbox{\raisebox{-1.ex}{$\stackrel
     {\textstyle>}{\textstyle\sim}$}}~7000$, the higher momenta begin to contribute to the 
growth of $\langle\delta \phi^2 \rangle$. 
Although we can expect the growth of low momentum modes
even for $x~\mbox{\raisebox{-1.ex}{$\stackrel
     {\textstyle>}{\textstyle\sim}$}}~7000$, it is rather inefficient and the fluctuation
produced at the final stage is mainly dominated by the higher momenta 
which are estimated by Eq.~$(\ref{S4})$.
The final variance is 
$\langle\delta \phi^2 \rangle_f=9.5 \times 10^{-8} M_{\rm PL}^2$ at 
$x_f =2.4 \times 10^4$.
The resonance stops by the back reaction effect of 
created particles and the final variance does not change so much 
compared with the $\xi=0$ case.
However, since $\tilde{\phi}_0 (x_f)=4.1 \times 10^{-4}$,
we can estimate as 
$\langle\delta \phi^2 \rangle_f=0.56 \tilde{\phi}_0^2$
at the termination point of the resonance.
This means that the energy transfer from the $\phi_0$
field to $\delta \phi_k$ particles becomes more efficient
with the increase of $|\xi|$.

We show in Fig.~7 the evolution of the fluctuation 
for $\xi=-70$ and $-80$ cases.  
The structure of the resonance is qualitatively same as the
$\xi = -50$ case.
Since the $\xi$ effect to the fluctuation continues 
longer with the increase of $|\xi|$,
the fluctuation reaches the first plateau at $x=7000$ and
$8000$ for $\xi=-70$ and $-80$ cases, respectively.
After a short plateau, the secondary resonant stage turns on and
$\langle\delta \phi^2 \rangle$ begins to increase.
However, at this stage, the growth rate is not so large as the first 
resonant stage, because the enhanced $k$ modes are restricted 
in the narrow range.  
We have found that the final variance slowly decreases with 
the decrease of $\xi$ for $-80~\mbox{\raisebox{-1.ex}{$\stackrel
     {\textstyle<}{\textstyle \sim}$}}~\xi~\mbox{\raisebox{-1.ex}{$\stackrel
     {\textstyle<}{\textstyle \sim}$}}~-50$.
For example, when $\xi=-50, -70$, and $-80$,
the final variances are $\langle\delta \phi^2
\rangle_f/M_{\rm PL}^2=9.5 \times
10^{-8}$, $7.8 \times 10^{-8}$ and $4.7 \times 10^{-8}$, respectively.
As for the ratio of $\langle\delta \phi^2 \rangle_f$ to $\tilde{\phi}_0^2$
at the maximum point of the fluctuation, 
$\langle\delta \phi^2 \rangle_f/\tilde{\phi}_0^2=0.56, 1.15, 1.16$ for 
$\xi=-50, -70, -80$, respectively.
Numerical calculations confirm that resonance stops when 
the fluctuation grows up to $\langle\delta \phi^2 \rangle_f \approx
\tilde{\phi}_0^2$ for $\xi~\mbox{\raisebox{-1.ex}{$\stackrel
     {\textstyle<}{\textstyle \sim}$}}~-60$.
Although $\tilde{\phi}_0 (x_f)$
becomes smaller as $\xi$ decreases, the energy of the oscillating inflaton 
field is transferred efficiently to the inflaton particles for 
$\xi~\mbox{\raisebox{-1.ex}{$\stackrel
     {\textstyle<}{\textstyle \sim}$}}~-60$
as compared with the minimally coupled case.
This is mainly due to the fact that nonadiabatic creation 
of the inflaton particles efficiently occurs around the minimum of 
the potential.

For $\xi~\mbox{\raisebox{-1.ex}{$\stackrel
     {\textstyle<}{\textstyle \sim}$}}~-100$, the structure of resonance is dominated
by the $\xi$ effect. 
We show the evolution of the inflaton fluctuation in Fig.~8
for the $\xi=-100$, $-200$ and $-1000$ cases.
We can not separate the resonant stage into two parts,
but find that low momentum modes are
enhanced continuously until $\langle\delta \phi^2 \rangle$
reaches to its maximum value.
Because of the nonexistence of the secondary inefficient resonant
stage, the resonance terminates rather early and 
$\tilde{\phi}_0 (x_f)$
takes larger value than that in the $-80~\mbox{\raisebox{-1.ex}{$\stackrel
     {\textstyle<}{\textstyle \sim}$}}~\xi~\mbox{\raisebox{-1.ex}{$\stackrel
     {\textstyle<}{\textstyle \sim}$}}~-50$ case.
For example, when $\xi=-200$, $\tilde{\phi}_0 \approx 1.0 \times 10^{-3}$,
which is much larger than the value 
$\tilde{\phi}_0 \approx 2.0 \times 10^{-4}$ in the $\xi=-80$ case.
This increase of the final value of $\tilde{\phi}_0$ results in 
the increase of the final fluctuation, since resonance ends by 
the back reaction effect when $\langle\delta \phi^2\rangle$ 
grows up to the comparable value with $\tilde{\phi}_0^2$.
For example, the final values are $\langle\delta \phi^2\rangle_f
=2.35 \times 10^{-7} M_{\rm PL}^2$ for $\xi=-100$;
$\langle\delta \phi^2 \rangle_f =1.86 \times 10^{-6} M_{\rm PL}^2$
for $\xi=-200$.
For $\xi~\mbox{\raisebox{-1.ex}{$\stackrel
     {\textstyle<}{\textstyle \sim}$}}~-200$, however, since the final value of 
$\tilde{\phi}_0$ becomes smaller with the decrease of $\xi$ again
due to the decrease of the initial value of $\phi_0$, the final 
variance also decreases. 
For example, $\langle\delta \phi^2\rangle=1.03 \times 
10^{-6} M_{\rm PL}^2$ for $\xi=-500$, 
and $\langle\delta \phi^2\rangle
=5.27 \times 10^{-7} M_{\rm PL}^2$ for $\xi=-1000$.
In the case of $\xi~\mbox{\raisebox{-1.ex}{$\stackrel
     {\textstyle<}{\textstyle \sim}$}}~-200$ the nonadiabatic change of 
the fluctuation in the vicinity of $\phi_0=0$ is clearly seen  
in Fig.~8.
This is because the  $\xi R a^2$ term in Eq.~$(\ref{B60})$
rapidly changes around $\phi_0=0$ especially when $|\xi|$ is 
very large as $|\xi|~\mbox{\raisebox{-1.ex}{$\stackrel
     {\textstyle>}{\textstyle\sim}$}}~200$.
However, in this case also, the growth of the fluctuation is finally
suppressed by the back reaction effect of created particles.

We have found that the final variance of the nonminimally coupled case
does not change significantly compared with the minimally coupled
case as is found in Fig.~9, where
the maximum value of $\langle\delta \phi^2\rangle$ is
$\langle\delta \phi^2\rangle_f =2 \times 10^{-6} M_{\rm PL}^2$ 
at $\xi \approx -200$.
However, the dispersion of the final fluctuation becomes comparable 
to the amplitude of the $\phi_0$ field for $\xi~\mbox{\raisebox{-1.ex}{$\stackrel
     {\textstyle<}{\textstyle \sim}$}}~-60$ (See Fig.~10).
This implies that although the initial value of inflaton
decreases as
increase of $|\xi|$, this is compensated by the efficient resonance 
with the appearance of new resonant modes close to $k=0$, 
We depict in Fig.~11 the distribution of produced momentum modes for 
the $\xi=0$, $-1$, $-50$, $-100$ cases.
With the decrease of $\xi$, we find that momentum modes 
close to $k=0$ are mainly enhanced. When $\xi~\mbox{\raisebox{-1.ex}{$\stackrel
     {\textstyle<}{\textstyle \sim}$}}~-100$, the final
fluctuation is dominated by low momentum modes, and this is 
different from the minimally coupled case.
As we have seen, preheating stage generically exists even if 
the nonminimal coupling is taken into account.

Finally, we should mention the case when $\lambda$
is changed. For largely negative $\xi$, larger values of 
$\lambda$  is allowed to fit the density perturbation
because the effective potential $(\ref{B50})$ becomes
flat with the increase of $|\xi|$.
As is found by Eq.~$(\ref{S21})$, the frequency of the $\phi_0$
field becomes large with the increase of $\lambda$,
which results in the fast growth of the fluctuation.
However, $\lambda$ is scaled out in the 
normalized equations of the $\phi_0$ and  $\delta\phi_k$ fields.
$\lambda$ appears only in the regularization of 
the initial amplitude of the fluctuation.
Since the growth of the fluctuation stops by the back reaction
effect of created particles, the final variance is almost the same
as the case of the fixed $\lambda$, as was suggested in the 
minimally coupled case in Ref.~\cite{KT3}.

\section{Concluding Remarks and Discussions}    
In this paper, we have investigated preheating of the nonminimally 
coupled inflaton field in the chaotic inflation model.
As for the massive inflaton case, since $\xi$ is constrained to be
small as $|\xi|~\mbox{\raisebox{-1.ex}{$\stackrel
     {\textstyle<}{\textstyle \sim}$}}~10^{-3}$ to lead a sufficient inflation, the
evolution of the system is almost the same as the minimally coupled case. 
In the self-coupling inflaton case, the constraint of $\xi$
is absent for negative $\xi$.
Hence we considered mainly this case and investigated how the 
ordinary picture of preheating is modified by taking into account
the $\xi$ effect.

In the minimally coupled case,
the equation of the inflaton fluctuation is 
reduced to the Lam\'e equation at the linear stage of preheating.
Although this model is not so efficient for the development
of the resonance compared with the massive inflaton plus another 
scalar field $\chi$ coupled to inflaton, the fluctuation 
$\langle\delta\phi^2\rangle$ increases by parametric resonance
and reaches to its maximum value $\langle\delta\phi^2\rangle
=1.5 \times 10^{-7} M_{\rm PL}^2$.
In reheating phase, the time-averaged scalar curvature vanishes
in this model, and the investigation of preheating is reduced to
the theory of Minkowski spacetime. 
Since the instability band is narrow as $3 \tilde{f}^2/2<
p^2<\sqrt{3} \tilde{f}^2$, the resonance is sensitive to the
back reaction effect of created particles.
The growth of the fluctuation ends at 
$\langle \delta \phi^2 \rangle_f \approx 0.05 \tilde{\phi}_0^2$, 
which means that
the dispersion of the fluctuation is only 20\% of the amplitude 
of the homogeneous inflaton field.

Taking into account the nonminimal coupling with the scalar 
curvature, the dynamics of preheating can be altered.  
We found that the $\xi$ effect works in two ways. 
First, since the initial value of inflaton for preheating 
becomes lowered as $|\xi|$ increases,
and the frequency of the oscillation of the inflaton field is reduced.
Hence the growth of the inflaton fluctuation is delayed by this effect
for fixed $\lambda$.
Second, since frequencies of the both homogeneous and fluctuational parts
change due to the nonminimal coupling, the structure of resonance becomes
different from the minimally coupled case. 
In the case of $-0.1~\mbox{\raisebox{-1.ex}{$\stackrel
     {\textstyle<}{\textstyle \sim}$}}~\xi<0$, the evolution of the 
system is almost the same as the minimally coupled case.
In  $-10~\mbox{\raisebox{-1.ex}{$\stackrel
     {\textstyle<}{\textstyle \sim}$}}~\xi~\mbox{\raisebox{-1.ex}{$\stackrel
     {\textstyle<}{\textstyle \sim}$}}~-0.1$ case, the decrease of 
the initial value of inflaton  mainly delays the evolution of the system.
Although the $\xi$ effect changes both frequencies of 
the $\phi_0$ and $\delta \phi_k$ fields at the initial 
stage of preheating, this works only for a short period.
When the particle creation rate surpasses 
the expansion rate of the Universe and 
$\langle\delta\phi^2\rangle$ begins to increase, 
the $\xi$ effect on the $\phi_0$ and $\delta \phi_k$ fields can be 
negligible. 
Hence, after that, the system of the evolution is not affected by the 
$\xi$ effect. 
The structure of resonance seems almost the same as 
the minimally coupled case, and produced
momentum modes are restricted in a narrow range
when $\langle\delta \phi^2\rangle$ reaches to its 
maximum value. The final variance does not change significantly
compared with the minimally coupled case.

As $\xi$ decreases further ($\xi~\mbox{\raisebox{-1.ex}{$\stackrel
     {\textstyle<}{\textstyle \sim}$}}~-10$), the initial stage of preheating 
is dominated by the effect of the nonminimal coupling.
The structure of resonance becomes different from that of the
$\xi=0$ case, and the wider range of momenta is allowed.
Particle creation occurs nonadiabatically in the vicinity of
$\phi_0=0$. 
Although the minimally coupled case is the least favorable one
for the development of the resonance because of the limited
resonant band, such limitation can be removed  
by taking into account the nonminimal coupling.
When $-80~\mbox{\raisebox{-1.ex}{$\stackrel
     {\textstyle<}{\textstyle \sim}$}}~\xi~\mbox{\raisebox{-1.ex}{$\stackrel
     {\textstyle<}{\textstyle \sim}$}}~-10$,  the stage of parametric resonance
is separated into two stages. At the first stage, the resonance
caused by the 
nonminimal coupling dominates, and 
momentum modes close to $k=0$ are efficiently enhanced.
With the passage of time, the $\xi$ effect onto equations of 
the $\phi_0$ and $\delta \phi_k$ fields becomes 
less important and the system 
enters the secondary resonant stage.
The second stage is the ordinary resonance stage of the minimally 
coupled case, and the growth rate of the fluctuation is not so
large as that of the first stage.
It is true that low momentum modes are enhanced, but
the final fluctuation is dominated by higher modes described 
by Eq.~$(\ref{S4})$.
Although the final variance slowly decreases with the decrease of 
$\xi$ for $-80~\mbox{\raisebox{-1.ex}{$\stackrel
     {\textstyle<}{\textstyle \sim}$}}~\xi~\mbox{\raisebox{-1.ex}{$\stackrel
     {\textstyle<}{\textstyle \sim}$}}~-10$, the ratio of the final fluctuation 
$\langle\delta \phi^2\rangle_f$ to the square of the amplitude of 
the $\phi_0$ field gets larger.
Especially for $\xi~\mbox{\raisebox{-1.ex}{$\stackrel
     {\textstyle<}{\textstyle \sim}$}}~-60$, we find 
$\langle\delta \phi^2\rangle_f \approx \tilde{\phi}_0^2$, which implies 
that the transfer of the energy
from the $\phi_0$ field to the fluctuation
is much more efficient than the minimally coupled case.

When $\xi~\mbox{\raisebox{-1.ex}{$\stackrel
     {\textstyle<}{\textstyle \sim}$}}~-100$, the growth of fluctuation ends before the 
secondary resonant stage appears, and parametric resonance is quite efficient.
For $-200~\mbox{\raisebox{-1.ex}{$\stackrel
     {\textstyle<}{\textstyle \sim}$}}~\xi~\mbox{\raisebox{-1.ex}{$\stackrel
     {\textstyle<}{\textstyle \sim}$}}~-100$, the final variance increases with the 
decrease of $\xi$ and $\langle\delta \phi^2\rangle_f$ takes the
maximum value $\langle\delta \phi^2\rangle_f =2 \times 10^{-6}
M_{\rm PL}^2$ at $\xi \approx -200$.  For $\xi~\mbox{\raisebox{-1.ex}{$\stackrel
     {\textstyle<}{\textstyle \sim}$}}~-200$, however, 
$\langle\delta \phi^2\rangle_f$ decreases because the initial
value of inflaton becomes smaller with the decrease of $\xi$, which 
yields the smaller amplitude at the termination point of resonance. 
We also found the relation 
$\langle\delta \phi^2\rangle_f \approx \tilde{\phi}_0^2$ 
for $\xi~\mbox{\raisebox{-1.ex}{$\stackrel
     {\textstyle<}{\textstyle \sim}$}}~-100$, which means that resonance 
mainly terminates by the back reaction effect 
when the dispersion of the fluctuation
grows up to the comparable value of the amplitude of the $\phi_0$
field.

Finally we have to comment on some points.
In this paper, we have made heavy use of the Hartree approximation.
In taking into account the back reaction effect of created particles,
there are other approaches. One of them is to add the stochastic 
noise term originated by quantum fluctuation to the inflaton equation.
This is based on the closed time path formalism in the non-equilibrium 
quantum field theory\cite{stoc}, and several authors made use of this 
method in the context of preheating\cite{stocpre}.
It is expected that the growth rate of the fluctuation would be changed
by taking into account this effect.
Other approach is the lattice simulation.
Although the final fluctuation of our numerical calculation in the 
minimally coupled case coincides well with that of the full nonlinear
calculation  in Ref.~\cite{KT3}, 
the final variance may change in the nonminimally coupled case 
performed by the lattice simulation. 
Since the lattice simulation includes the rescattering effect which is 
absent in the Hartree approximation, the spectrum of enhanced momentum
modes would also change by rescattering of created particles.
Although we do not present details of the difference in this paper,
we should  consider preheating by various approaches of the
back reaction effect  and compare them with the fully nonlinear
calculation.

Recently, several authors considered the metric perturbation in reheating
phase~\cite{mper}. As for the self-coupling inflaton model, 
it was pointed out in Ref.~\cite{PE} that the metric perturbation 
is resonantly amplified even if we do not introduce another scalar field  coupled to inflaton.
In the present model of the nonminimal coupling, since the super Horizon 
modes can be enhanced by the strong nonminimal coupling, this may
alter the spectrum of the density perturbation.
It will be interesting to investigate preheating including the metric
perturbation, because modification of the spectrum of the density 
perturbation would result in some important consequences such as the 
production of primordial black holes.
These issues are under consideration.

\section*{ACKOWLEDGEMENTS}
S. T. would like to thank B. A. Bassett and A. Taruya for useful discussions.
T. T. is thankful for financial support from the JSPS. This
work was supported partially by a Grant-in-Aid for  Scientific
Research Fund of the Ministry of Education, Science and Culture
(No. 09410217 and Specially Promoted Research No. 08102010), by a
JSPS Grant-in-Aid (No. 094162), and by the Waseda University Grant 
for Special Research Projects.


\vskip 5cm

\begin{table}
\caption{The final value of $|\alpha|$ and $\phi_0$ when 
the inflationary period ends in the $\lambda\phi^4/4$ model 
for negative $\xi$. We find that $|\alpha_F|$ takes the almost constant value 
$|\alpha_F| \approx 1$ for $\xi <-1$.
On the other hand, $\phi_0(t_F)$ becomes smaller as $|\xi|$ increases.
}

\vskip .3cm
\noindent
\begin{tabular}{crc|ccc}
        ~& $\xi$ & ~&
         $|\alpha_F|$ &
         $\phi_0(t_F)/M_{\rm PL}$ 
         &~\\
        \hline
~&$0$ &~& 0 &0.5642 &~\\
~&$-0.1$ &~& 0.4606 &0.4281 &~\\
~&$-1$ &~& 1.0000 &0.1995 &~\\
~&$-10$ &~& 1.1370 &0.0673&~\\
~&$-50$ &~& 1.1511 &0.0303&~\\
~&$-100$ &~& 1.1529 &0.0215&~\\
~&$-1000$ &~& 1.1545 &0.0068&~\\
\end{tabular}
\end{table}

\vskip 2cm
\begin{flushleft}
{ Figure Captions}
\end{flushleft}
\noindent
\parbox[t]{2cm}{FIG. 1:\\~}\ \
\parbox[t]{8cm}
{The potential in the equivalent system to the nonminimally
coupled inflaton.
 The dotted  and solid curves denote
the massive inflaton and the massless inflaton case 
with $\xi=-0.1$ respectively.
The potential of the massive inflaton has a local maximum and
inflation is difficult to realize unless $|\xi|~\mbox{\raisebox{-1.ex}{$\stackrel
     {\textstyle<}{\textstyle \sim}$}}~10^{-3}$. 
However, the potential of the massless inflaton
 has a flat plateau  to lead a sufficient inflation for negative $\xi$.
}\\[1em]
\noindent
\parbox[t]{2cm}{FIG. 2:\\~}\ \
\parbox[t]{8cm}
{The evolution of $\langle \delta \bar{\phi}^2 \rangle$ as a function of
$x$ in $\xi=0$ and $\xi=-1$ cases. In both cases, the fluctuation grows
by parametric resonance. When $\xi=-1$, since the initial value of inflaton 
becomes smaller compared with the $\xi=0$ case,
this makes the growth of the fluctuation delayed.
However, the final variance $\langle \delta \bar{\phi}^2 \rangle
\approx 10^{-7}$ is almost the same as the minimally coupled case. 
 }\\[1em]
\noindent
\parbox[t]{2cm}{FIG. 3:\\~}\ \
\parbox[t]{8cm}
{The evolution of the homogeneous inflaton field as a function of
$x$ in $\xi=0$ and $\xi=-1$ cases.
Since the initial value of inflaton for the $\xi=-1$ case is 
smaller than that of the $\xi=0$ case, the time interval of the 
oscillation becomes longer.
}\\[1em]
\noindent
\parbox[t]{2cm}{FIG. 4:\\~}\ \
\parbox[t]{8cm}
{The evolution of the $\phi_0$ field as a function of $x$ in the case of 
$\xi=-100$. The $\phi_0$ field slowly changes around the local
 maximum point of the potential, but nonadiabatically changes
 in the vicinity of $\phi_0 \approx 0$.
}\\[1em]
\noindent
\parbox[t]{2cm}{FIG. 5:\\~}\ \
\parbox[t]{8cm}
{The evolution of $\omega_1 \equiv 3f^2 \left(1+
\frac{\langle \delta \bar{\varphi}^2 \rangle}
{f^2} \right)$ and 
$\omega_2 \equiv \left(\xi-\frac16 \right) a^2 \bar{R}$ 
as a function of $x$ in the case of $\xi=-20$.
The effect by the $\omega_2$ term is significant at the 
initial stage of preheating, but it gradually becomes negligible
with the passage of time. 
}\\[1em]
\noindent
\parbox[t]{2cm}{FIG. 6:\\~}\ \
\parbox[t]{8cm}
{The evolution of $\langle \delta \bar{\phi}^2 \rangle$ as a function of
$x$ in $\xi=-20$ and $\xi=-50$ cases. At the initial stage of preheating,
the structure of resonance is modified and the range of enhanced momenta 
becomes wider. However, this stage ends before 
$\langle \delta \bar{\phi}^2 \rangle$ sufficiently increases
and the second resonance stage sets in.
In both cases of $\xi=-20$ and $\xi=-50$, the growth rate of the
fluctuation becomes smaller after the first resonance stage ends.
}\\[1em]
\noindent
\parbox[t]{2cm}{FIG. 7:\\~}\ \
\parbox[t]{8cm}
{The evolution of $\langle \delta \bar{\phi}^2 \rangle$ as a function of
$x$ in $\xi=-70$ and $\xi=-80$ cases. We can separate the resonance 
into two stages. The first stage is dominated by the effect of the
nonminimal coupling, which connects to the second stage of ordinary minimal
coupling. With the decrease of $\xi$, the initial stage becomes longer.
}\\[1em]
\noindent
\parbox[t]{2cm}{FIG. 8:\\~}\ \
\parbox[t]{8cm}
{The evolution of $\langle \delta \bar{\phi}^2 \rangle$ as a function of
$x$ in $\xi=-100$, $\xi=-200$ and $\xi=-1000$ cases. 
In these cases, the fluctuation reaches to its maximum value only by the
first efficient resonance stage by the effect of the nonminimal 
coupling. Therefore, the final fluctuation is dominated by the momenta
close to $k=0$.
}\\[1em]
\noindent
\parbox[t]{2cm}{FIG. 9:\\~}\ \
\parbox[t]{8cm}
{The final value of $\langle \delta \bar{\phi}^2 \rangle$ 
as a function of $|\xi|$ for the negative $\xi$.
In the case of $-80~\mbox{\raisebox{-1.ex}{$\stackrel
     {\textstyle<}{\textstyle \sim}$}}~\xi~\mbox{\raisebox{-1.ex}{$\stackrel
     {\textstyle<}{\textstyle \sim}$}}~0$, 
$\langle \delta \bar{\phi}^2 \rangle_f$ slowly decreases with the 
increase of $|\xi|$. For $\xi~\mbox{\raisebox{-1.ex}{$\stackrel
     {\textstyle<}{\textstyle \sim}$}}~-100$, the resonance 
becomes efficient by the $\xi$ effect,
and $\langle \delta \bar{\phi}^2 \rangle_f$ takes the maximum value 
$\langle \delta \bar{\phi}^2 \rangle_f \approx 2 \times 10^{-6}$
at $\xi \approx -200$. 
However, $\langle \delta \bar{\phi}^2 \rangle_f$ slowly decreases for 
$\xi~\mbox{\raisebox{-1.ex}{$\stackrel
     {\textstyle<}{\textstyle \sim}$}}~-200$ because the growth of the fluctuation is suppressed
by the back reaction effect of created particles.
}\\[1em]
\noindent
\parbox[t]{2cm}{FIG. 10:\\~}\ \
\parbox[t]{8cm}
{The ratio of the final variance $\langle \delta \phi^2 \rangle_f$
to the square of the amplitude of the $\phi_0$ field.
This ratio increases with the decrease of $\xi$ for $-50~\mbox{\raisebox{-1.ex}{$\stackrel
     {\textstyle<}{\textstyle \sim}$}}~\xi~
\mbox{\raisebox{-1.ex}{$\stackrel
     {\textstyle<}{\textstyle \sim}$}}~0$, and reaches the plateau 
$\langle \delta \phi^2 \rangle_f/\tilde{\phi}_0^2 \approx 1$ 
for $\xi~\mbox{\raisebox{-1.ex}{$\stackrel
     {\textstyle<}{\textstyle \sim}$}}~-60$.
}\\[1em]
\noindent
\parbox[t]{2cm}{FIG. 11:\\~}\ \
\parbox[t]{8cm}
{The spectrum of enhanced momentum modes in $\xi=0$, $\xi=-1$,
 $\xi=-50$, and $\xi=-100$ cases. Although the resonance band is 
narrow in the minimally coupled case, the low momentum modes 
begin to be enhanced with the decrease of $\xi$.
When $\xi~\mbox{\raisebox{-1.ex}{$\stackrel
     {\textstyle<}{\textstyle \sim}$}}~-100$, the final fluctuation is dominated by low
momentum modes. 
}\\[1em]
\noindent


\begin{thebibliography}{99}
\bibitem{com}
In other papers, the value of $\lambda=10^{-13}$ is often adopted.
However, we have confirmed that the difference for these two cases of
$\lambda$ is negligibly small and hardly alters 
our results and discussions.
\bibitem{comment}
We have also obtained the approximate equation of the fluctuation
around $\phi_0 = \tilde{\phi}_0$.
This is expressed as $\frac{d^2}{dx^2} \delta \varphi_k
+\left( p^2 +2f^2 \right) 
\delta \varphi_k \approx 0$, and the structure of resonance is different from 
that of the minimally 
coupled case. In this case, however, since the $\phi_0$ field 
moves very slowly, the nonadiabatic increase of 
the fluctuation hardly occurs.
\end{thebibliography}
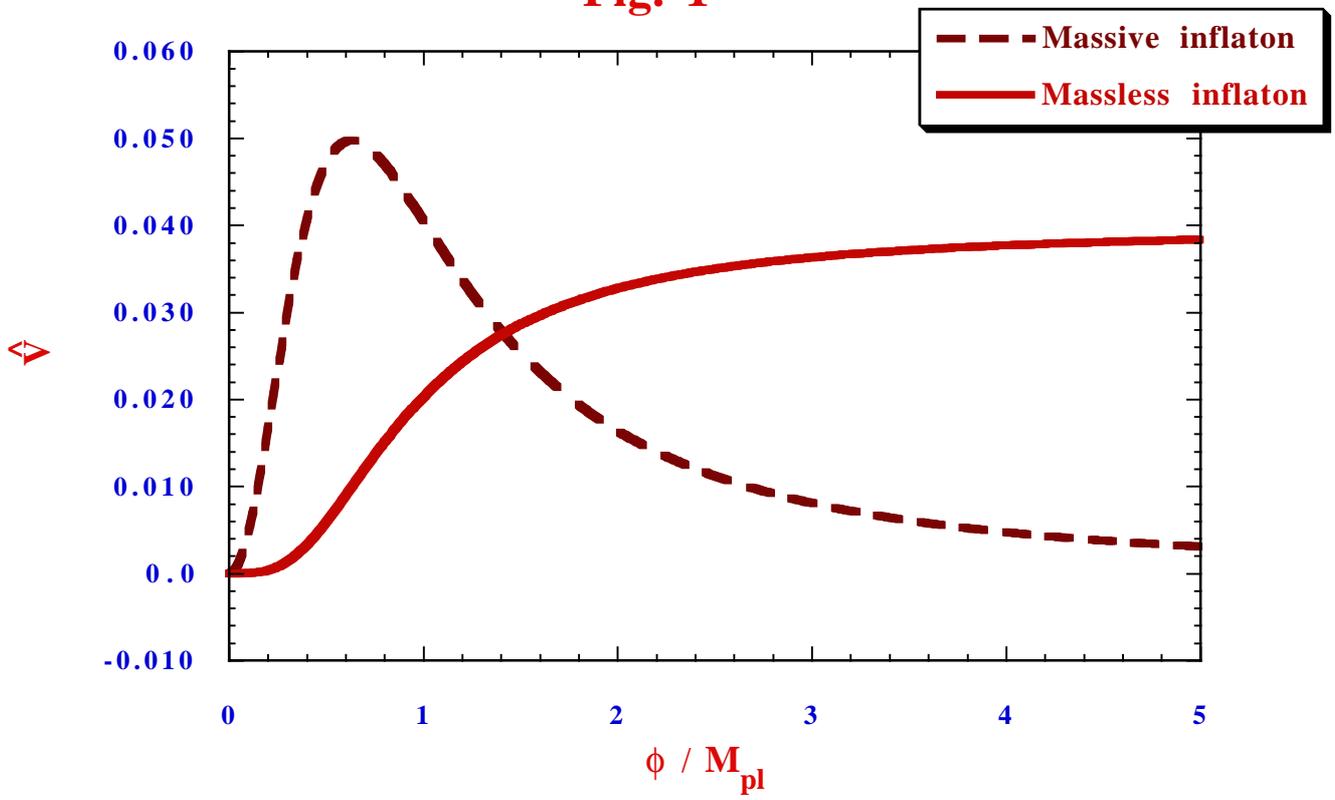
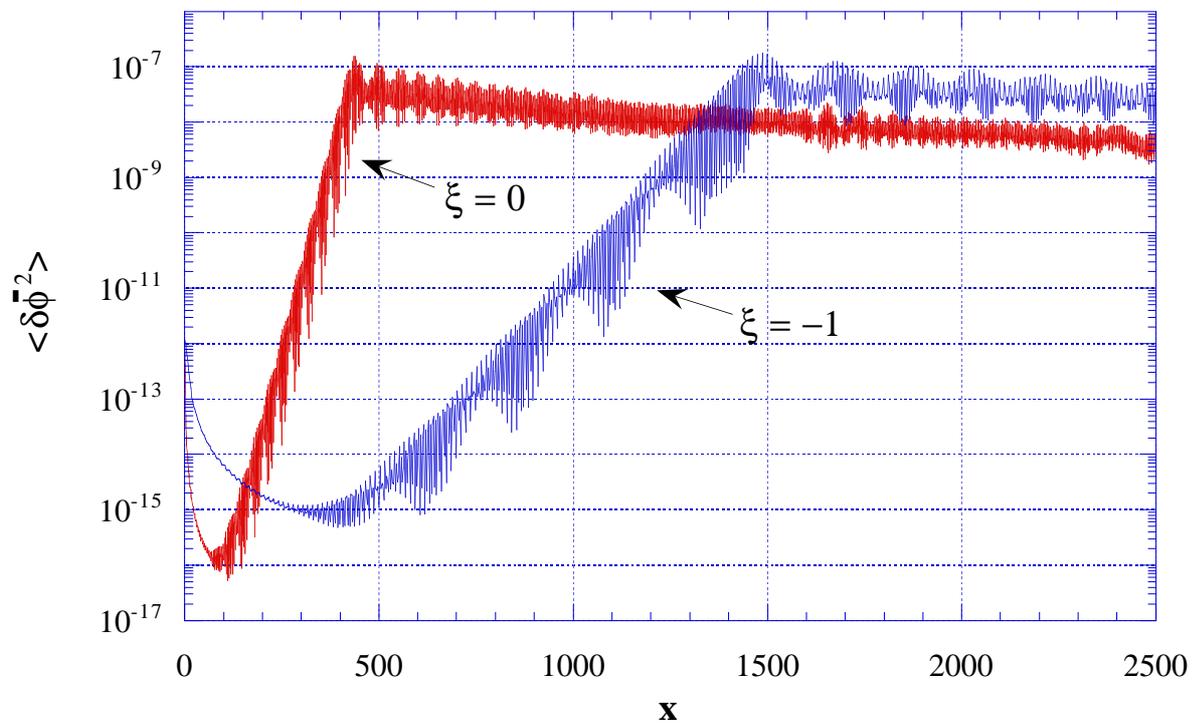
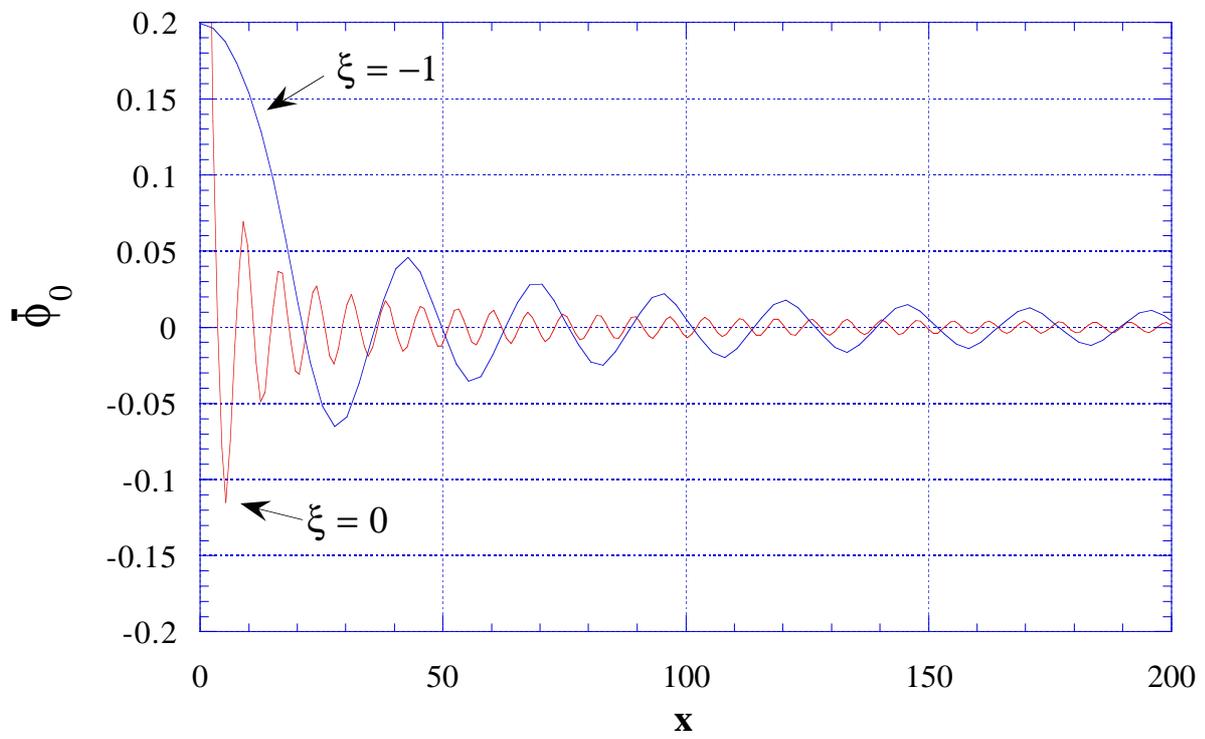
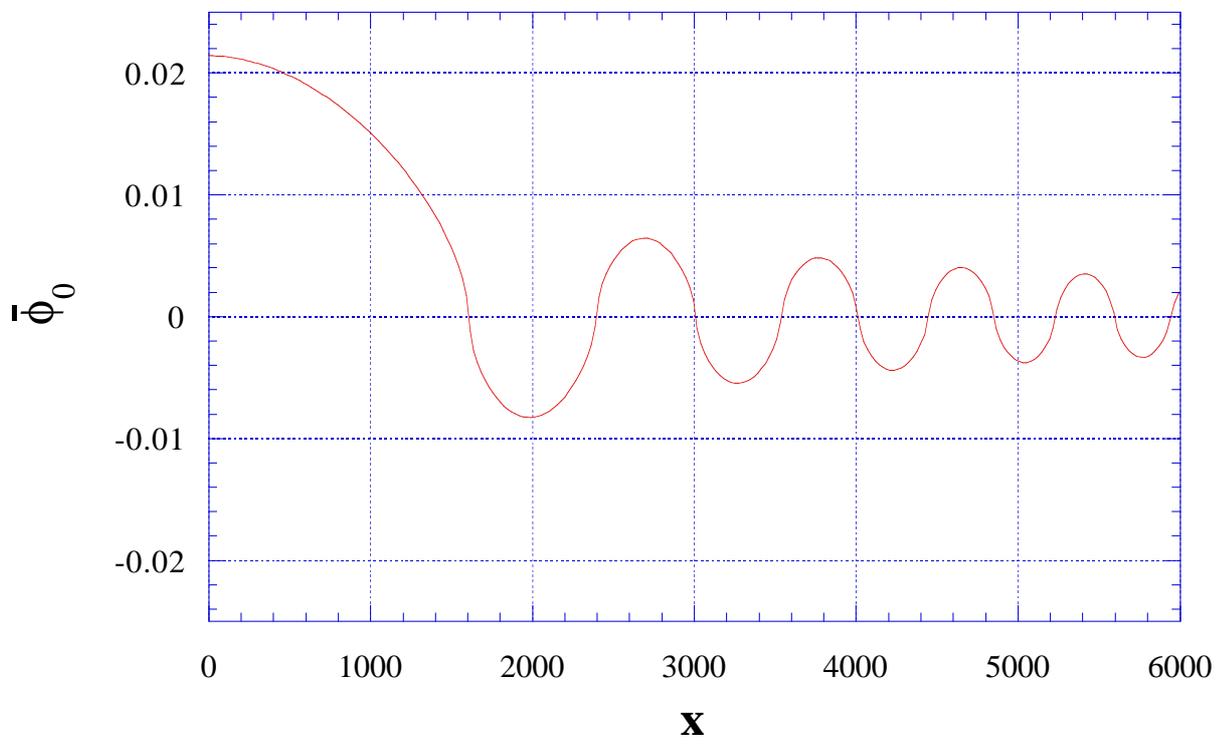
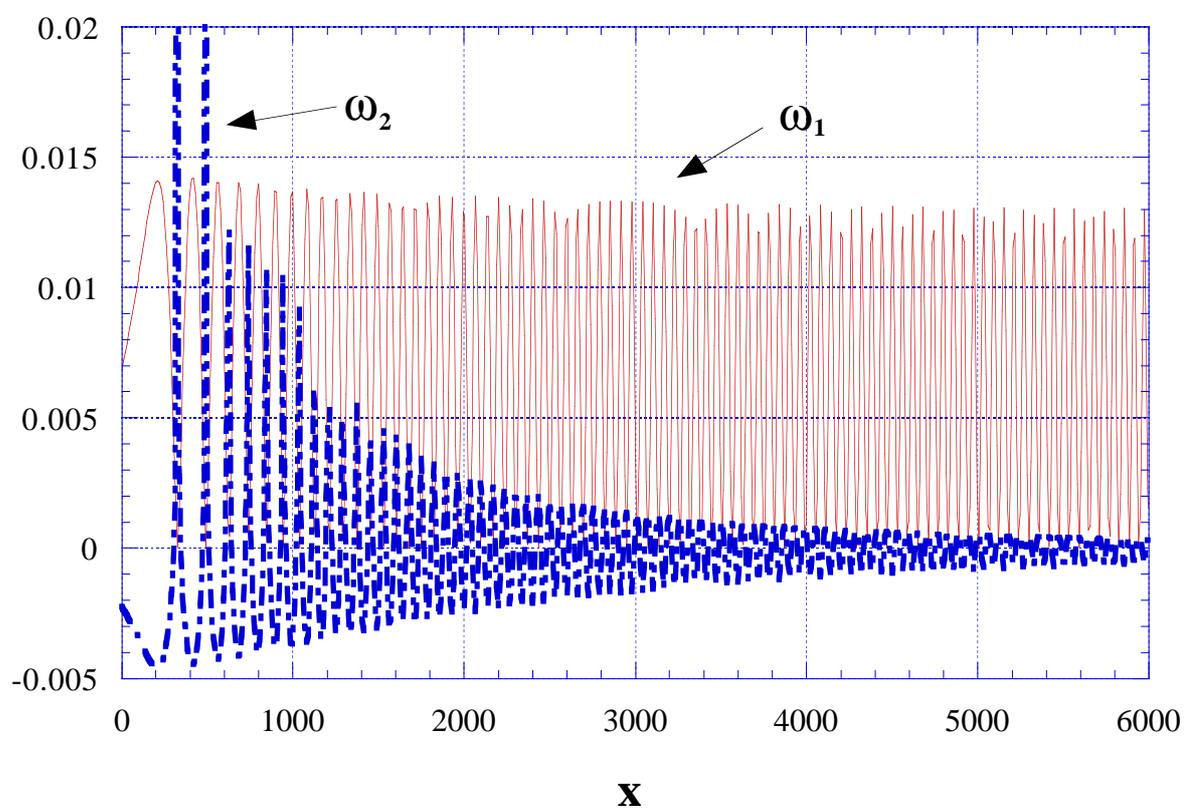
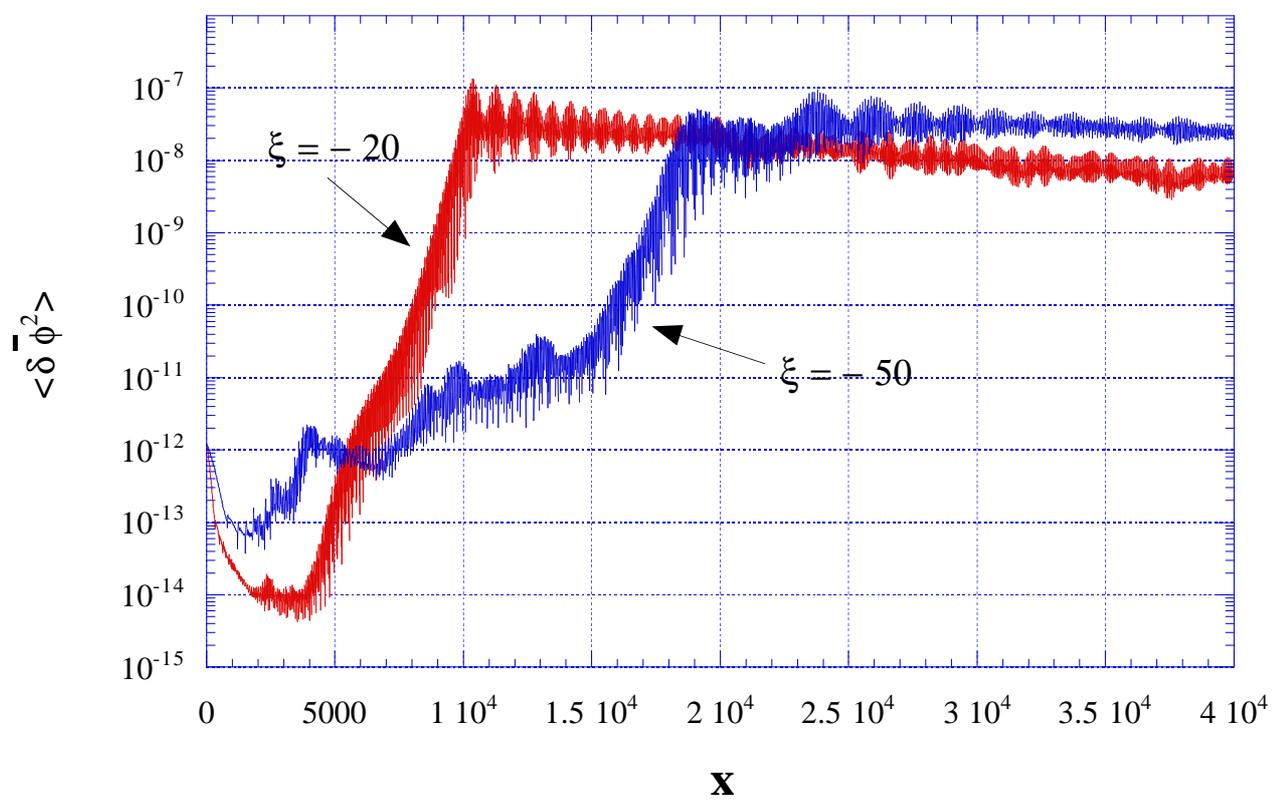
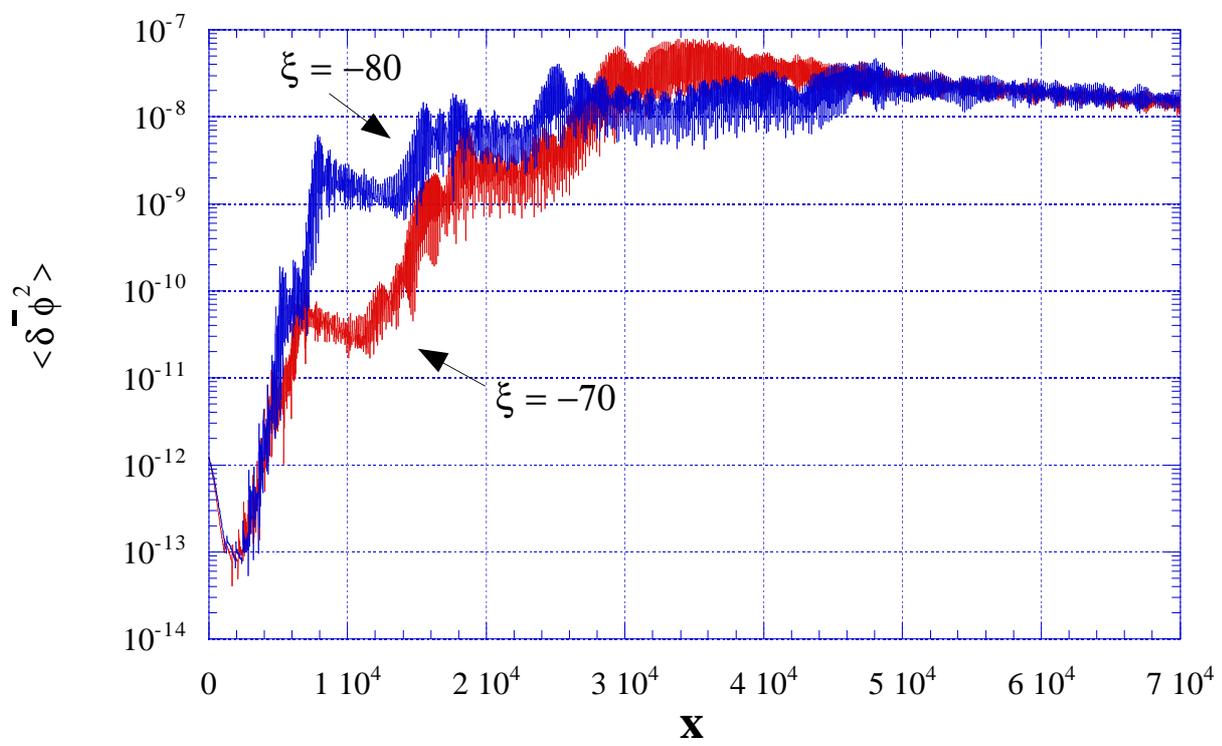
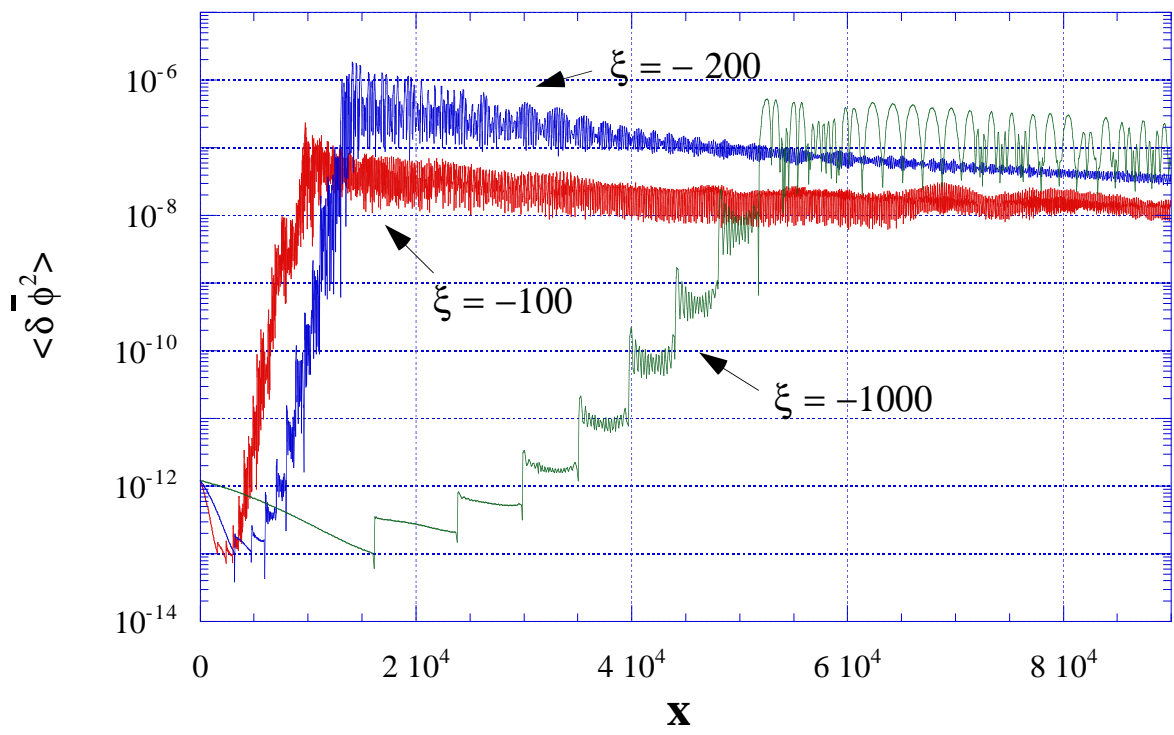
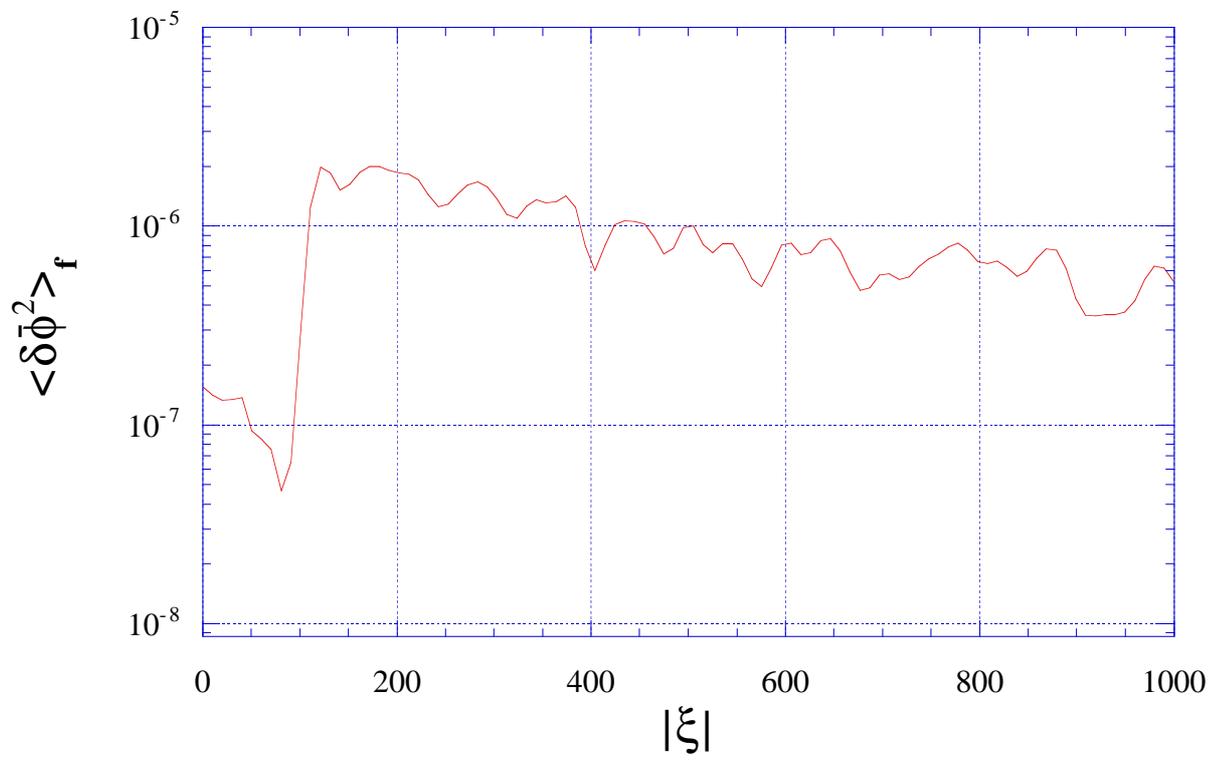
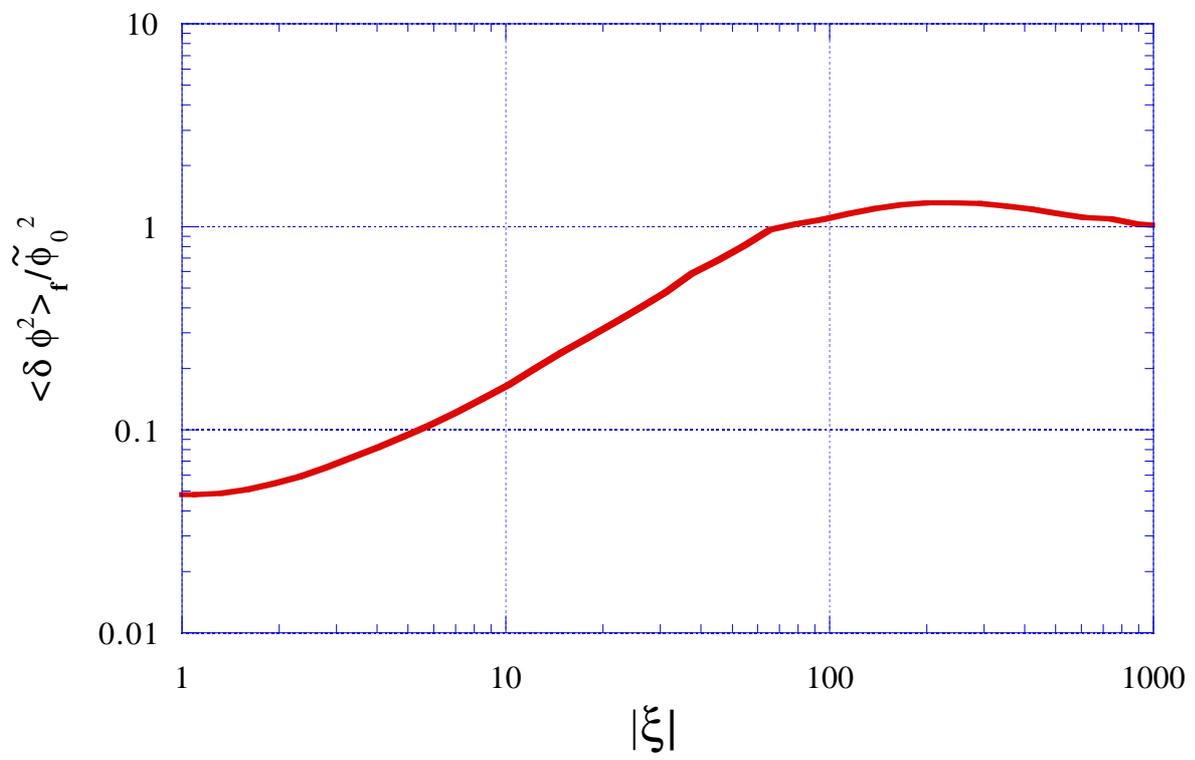
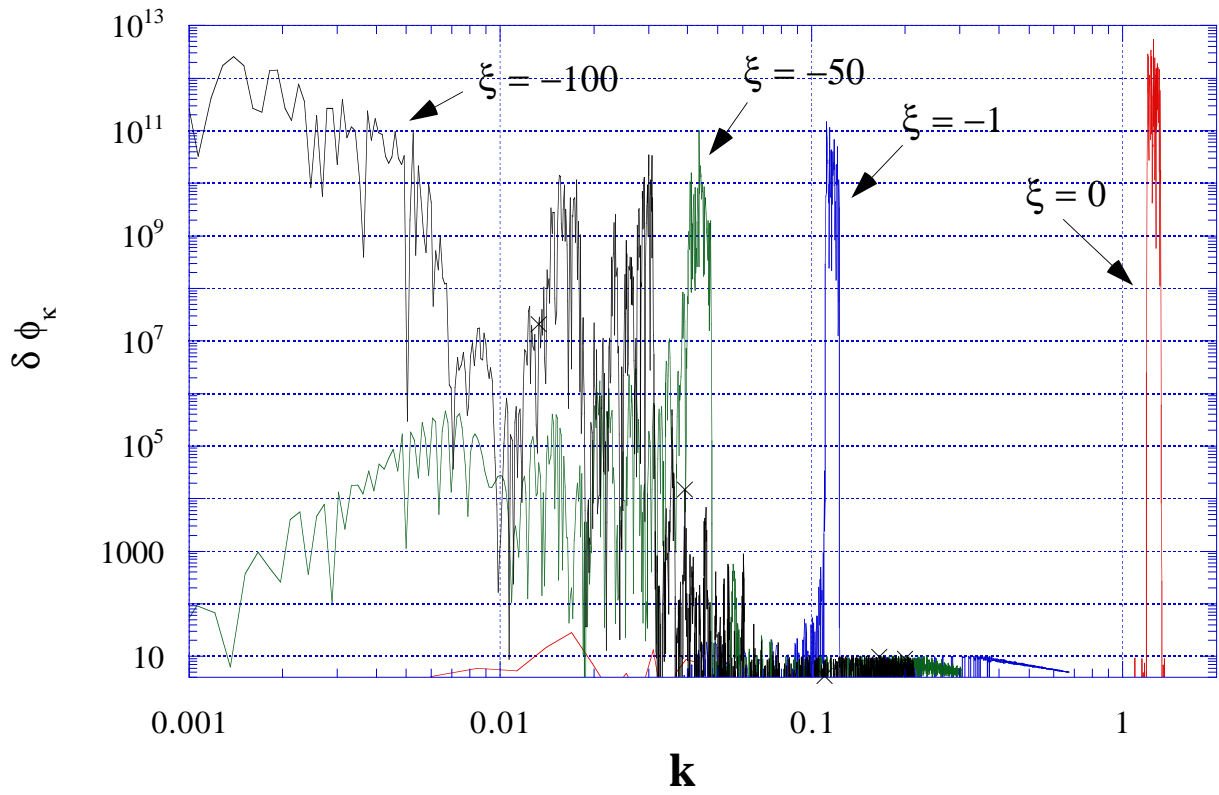
\end{document}